%% file: jsac-open-ran-paper-2023-v3.tex
\documentclass[10pt, journal, letterpaper]{IEEEtran}
\usepackage{url}
\usepackage[utf8]{inputenc}
\usepackage[dvipsnames]{xcolor}
\usepackage{amsmath}
\usepackage{amssymb}
\usepackage{xspace}
\usepackage{epsfig}
\usepackage{balance}

\usepackage[acronyms,nonumberlist,nopostdot,nomain,nogroupskip]{glossaries}
\usepackage{booktabs}
\usepackage{tabularx}
\usepackage{makecell}
\usepackage{tikz}
\usepackage{pgfplots}
\pgfplotsset{compat=newest}
\pgfplotsset{plot coordinates/math parser=false}
\newlength\fheight
\newlength\fwidth
\usetikzlibrary{plotmarks,patterns,decorations.pathreplacing,backgrounds,calc,arrows,arrows.meta,spy,matrix,scopes}
\usepgfplotslibrary{patchplots,groupplots}
\usepackage{tikzscale}
\usepackage[draft]{hyperref}

\newif\ifexttikz
\exttikzfalse
\ifexttikz
	\usetikzlibrary{external}
	\usepackage{fontspec}
\fi

\usepackage[var0]{inconsolata}

\usepackage{multirow}

\usepackage[font=footnotesize]{subcaption}
\usepackage[font=footnotesize]{caption}

\usepackage{mathtools}
\usepackage[numbers,sort&compress]{natbib}

\usepackage{soul}


\input{acronyms.tex}
\input{myTikz.tex}

\usepackage{dblfloatfix}

\makeglossaries

\usepackage{listings}

\definecolor{codegray}{rgb}{0.25,0.25,0.25}
\definecolor{codepurple}{rgb}{0.58,0,0.82}
\lstdefinestyle{mystyle}{
  commentstyle=\color{PineGreen},
  keywordstyle=\color{MidnightBlue}\bfseries,
  numberstyle=\tiny\color{codegray},
  stringstyle=\color{codepurple},
  basicstyle=\ttfamily\scriptsize,
  breakatwhitespace=true,         
  breaklines=true,                 
  captionpos=b,
  frame=tb,
  keepspaces=true,                 
  numbers=left,                    
  numbersep=5pt,                  
  showspaces=false,                
  showstringspaces=false,
  showtabs=false,                  
  tabsize=2,
  xleftmargin=10pt,
  belowskip=-10pt,
  float=htbp,  % prevents listing from being split across multiple pages
}
 
\lstdefinelanguage{myxml}{%
  language     = XML,
  morekeywords = {E2AP,PDU,criticality,value,id},
}

\usepackage{xspace}

\newcommand{\ran}{\gls{ran}\xspace}

\newcommand{\nearrt}{\gls{near-rt}\xspace}
\newcommand{\nonrt}{\gls{non-rt}\xspace}

\newcommand{\ric}{\gls{ric}\xspace}
\newcommand{\rics}{\glspl{ric}\xspace}

\begin{document}

\title{Empowering the 6G Cellular Architecture\\with Open RAN}

\author{
% \IEEEauthorblockN{JSAC GEs}
\IEEEauthorblockN{Michele Polese, Mischa Dohler, Falko Dressler, Melike Erol-Kantarci,\\Rittwik Jana, Raymond Knopp, Tommaso Melodia}
\thanks{M. Polese and T. Melodia are with the Institute for the Wireless Internet of Things, Northeastern University, Boston, MA, USA. E-mail: \{m.polese, melodia\}@northeastern.edu.}
\thanks{M. Dohler is with Advanced Technology Group, Ericsson Inc., Santa Clara, CA, USA. E-mail: mischa.dohler@ericsson.com.}
\thanks{F. Dressler is with the School of Electrical Engineering and Computer Science, TU Berlin, Germany. Email: dressler@ccs-labs.org.}
\thanks{M. Erol-Kantarci is with the School of Electrical Engineering and Computer Science, University of Ottawa, Ottawa, ON, Canada. E-mail: melike.erolkantarci@uottawa.ca.}
\thanks{R. Jana is with Google, NYC, NY. E-mail: rittwikj@google.com.}
\thanks{R. Knopp is with Eurecom, Sophia Antipolis. E-mail: raymond.knopp@eurecom.fr.}

\thanks{This work was partially supported by the U.S.\ National Science Foundation under Grant CNS-2117814 as well as by the Federal Ministry of Education and Research (BMBF, Germany) within the 6G Platform under Grant 16KISK050.}
}

% decrease space after author block
% \makeatletter
% \patchcmd{\@maketitle}
%   {\addvspace{0.5\baselineskip}\egroup}
%   {\addvspace{-1.5\baselineskip}\egroup}
%   {}
%   {}
% \makeatother

\flushbottom
\setlength{\parskip}{0ex plus0.1ex}

\maketitle
\glsunset{nr}
\glsunset{lte}
\glsunset{3gpp}
\glsunset{usrp}
\glsunset{sctp}

\begin{abstract}
Innovation and standardization in 5G have brought advancements to every facet of the cellular architecture. This ranges from the introduction of new frequency bands and signaling technologies for the \gls{ran}, to a core network underpinned by micro-services and \gls{nfv}. 
However, like any emerging technology, the pace of real-world deployments does not instantly match the pace of innovation. To address this discrepancy, one of the key aspects under continuous development is the \gls{ran} with the aim of making it more open, adaptive, functional, and easy to manage.

In this paper, we highlight the transformative potential of embracing \emph{novel cellular architectures} by transitioning from conventional systems to the progressive principles of Open \gls{ran}. This promises to make 6G networks more agile, cost-effective, energy-efficient, and resilient. It opens up a plethora of novel use cases, ranging from ubiquitous support for autonomous devices to cost-effective expansions in regions previously underserved.
The principles of Open \gls{ran} encompass: (i) a disaggregated architecture with modular and standardized interfaces; (ii) cloudification, programmability and orchestration; and (iii) AI-enabled data-centric closed-loop control and automation.
We first discuss the transformative role Open RAN principles have played in the 5G era.  Then, we adopt a system-level approach and describe how these Open \gls{ran} principles will support 6G \gls{ran} and architecture innovation. We qualitatively discuss potential performance gains that Open \gls{ran} principles yield for specific 6G use cases. For each principle, we outline the steps that research, development and standardization communities ought to take to make Open \gls{ran} principles central to  next-generation cellular network designs. 
\end{abstract}


\begin{picture}(0,0)(10,-505)
\put(0,0){
\put(0,0){\small This paper is part of the IEEE JSAC Special Issue on Open RAN. DOI: 10.1109/JSAC.2023.3334610}
\put(0,-10){
\footnotesize  \copyright IEEE 2023. Personal use of this material is permitted. Permission from IEEE must be obtained for all other uses, in any current or future media} 
\put(0,-20){
\footnotesize  including reprinting/republishing
this material for advertising or promotional purposes, creating new collective works, for resale or redistribution to servers} 
\put(0,-30){
\footnotesize  or lists, or reuse of any copyrighted component of this work in other works.}}
\end{picture}

\glsresetall
\glsunset{nr}
% \glsunset{lte}
\glsunset{3gpp}
\glsunset{usrp}
\glsunset{sctp}
\glsunset{4g}
\glsunset{5g}

\section{Introduction}
\label{sec:intro}

\glsunset{5g}

The wireless internet plays a fundamental role today in supporting societies and economies around the world,  underpinned by cellular connectivity that is widely used by consumers but also in industries, health, education, and entertainment. This is testament to the revolution that fourth and fifth generations (4G and 5G) of cellular networks have introduced, making it easy to stream data at high rates to mobile phones, to provide connectivity to vehicles and sensors, and so much more. 

More than ten years of releases from \gls{3gpp} have defined the \gls{lte} and \gls{nr} technologies underpinning the 4G and 5G \gls{ran} and core network~\cite{parkvall2017nr}. These cellular systems share an IP-based core network design, an \gls{ofdm} waveform in the \gls{ran} physical layer, end-to-end data abstractions to apply security and \gls{qos} to user flows, and robust mobility procedures, among many other novel principles. \gls{5g} further advances cellular systems with an array of first-of-its-kind solutions, including support for communications in the lower portion of the \gls{mmwave} band~\cite{pi2011intro}, directional communications~\cite{rangan2017potentials}, massive \gls{mimo}~\cite{marzetta2010noncooperative}, and a frame structure which supports different user traffic requirements~\cite{lien20175g}. 

% \hl{What is missing}
Nonetheless, while the standard specifications feature several relevant techniques to provide data rates, latency, and other \glspl{kpm} in line with the \gls{5g} definitions from the \gls{itu}~\cite{itu-r-2083}, actual \gls{5g} deployments lag behind the standardization by at least one release: indeed, the \gls{5g} Releases 16 and 17 have already been finalized whilst only a subset of Release 16 features is commonly deployed, and often in non-standalone mode~\cite{list5gnetwork}. In addition, the usage of the \gls{mmwave} spectrum has not taken off~\cite{narayanan2022comparative}. Finally, today's 5G networks are largely deployed using global default configurations, neglecting the potential optimization benefits available with bespoke and tailored setups~\cite{bonati2021intelligence,mollahasani2021actor}.

This is because of several reasons, however, financial being the main one. And whilst the largest bottom-line item on the balance sheet for cellular operator remains customer churn and the largest unrealized top-line item remains the inability to charge for innovative 5G use-cases, an important cost factor remains: the high deployment and operational costs that are associated to installing, configuring, optimizing, and operating \gls{5g} equipment without a clear business strategy for the return on investment. 
While the research and development ecosystem has limited control on the business and market development, we believe that there are technical, architectural, and system-level solutions that, if natively integrated in the design of \gls{6g} networks, can kickstart a faster innovation cycle for cellular, introducing approaches that are widely adopted in cloud environments and the software industry.

Discussions are currently well underway on what kind of technologies and use cases are to be expected for 6G~\cite{yang20196g,giordani2020toward,saad20206g,dang2020should,letaief2019roadmap,tataria2022six}. The recent literature, however, lacks a vision on how we can address fundamental issues related to the network architecture and introduce a clean-slate, system-level re-design of next-generation cellular networks. This paper fills this gap by providing a systematic review of the architectural limitations of existing cellular networks and deployments, and and a tutorial on how the foundational principles of Open \gls{ran} can be embedded into the \gls{6g} cellular architecture to address some of such limitations.

Notably, we envision \gls{6g} networks to be data-driven, autonomous systems where algorithmic control and \gls{ai} are systematically applied to fine-tune and optimize programmable protocol stacks. Further, virtualized (and possibly disaggregated) network functions are dynamically and automatically orchestrated to comply with specific performance requirements of use cases and application. These networks are primarily deployed through software, with hardware acceleration, and can make an agile use of dedicated or shared spectrum, compute, and infrastructure resources. They can be deployed with minimal effort, and can extend coverage through automated and optimzed self-backhaul solutions. Finally, Open-\gls{ran}-based 6G systems can leverage the same principles of programmability, automation, and virtualization to adopt cloud-native zero-trust security and resiliency strategies. The innovation introduced by Open \gls{ran} in wireless systems is comparable to the transformation that \gls{sdn} and programmable user planes have brought along in Ethernet switching systems~\cite{mckeown2008openflow,bosshart2014p4}, opening opportunities for closed-loop and intelligent algorithmic control of once inflexible platforms. 

The main contribution of this paper is to clearly articulate and describe the key constituents that are required to implement this vision, providing a tutorial that can illustrate and direct research efforts toward Open \gls{ran} in 6G. Compared to~\cite{polese2023understanding,bonati2020open,lee2020hosting,bonati2021intelligence,abdalla2021generation,garciasaavedra2021oran,brik2022deep,arnaz2022toward,app12010408,tataria2022six}, which mostly focus on the building blocks and the development of Open \gls{ran} systems, in this paper we analyze cellular networks as complex systems and identify opportunities towards innovation at the \emph{system level}. The second contribution of this paper is connecting key \gls{6g} performance requirements (e.g., energy efficiency, ubiquitous coverage, resiliency, low cost and complexity) to Open \gls{ran} principles, discussing the performance and operational gains that open, virtualized, and intelligent networks enable. Finally, for each of the building blocks of Open RAN systems, we discuss current state of the art and research directions.

Overall, this paper provides a map and a vision to steer research and development in Open \gls{ran} into 6G network, bringing system design to the forefront of the conversation on what 6G networks will be. Specifically, in Section~\ref{sec:trad-arch}, we revisit 4G and 5G legacy architectures. In Section~\ref{sec:new-arch}, we discuss the Open RAN principles. In Section~\ref{sec:program}, we deep-dive into important enablers that underpin Open RAN principles. We then focus on the three main tenants of Open RAN, i.e. disaggregation in Section~\ref{sec:disaggregation}, cloud nativeness in Section~\ref{sec:cloud} and closed-loop control in Section~\ref{sec:closed-loop}. We finish the article by discussing two emerging opportunities, notably dynamic spectrum sharing in Section~\ref{sec:sharing} and self-configurable joint access and backhaul in Section~\ref{sec:iab}. We conclude in Section~\ref{sec:conclusions}.

\section{4G and 5G Cellular Network Architecture}
\label{sec:trad-arch}

The design of cellular networks separates functionalities across different elements---the \gls{ran} and core network, in a horizontal split---and planes---the user and the control planes (UP and CP), in a vertical split~\cite{3gpp.38.300}. They support connectivity in \emph{mobile scenarios} and provide \emph{exclusive} spectrum access to licensed and shared bands through centralized scheduling. Compared to other popular wireless technologies such as Wi-Fi, cellular networks support a more widespread coverage layer with a consistent experience also in mobile scenario, and can extend performance guarantees to the end users. However, existing deployments still present limitations, e.g., they come with limited reconfigurability and adaptability to use cases and users' requirements and rely partly on manual optimization.

\begin{figure}[t]
    \centering
    \includegraphics[width=.8\linewidth]{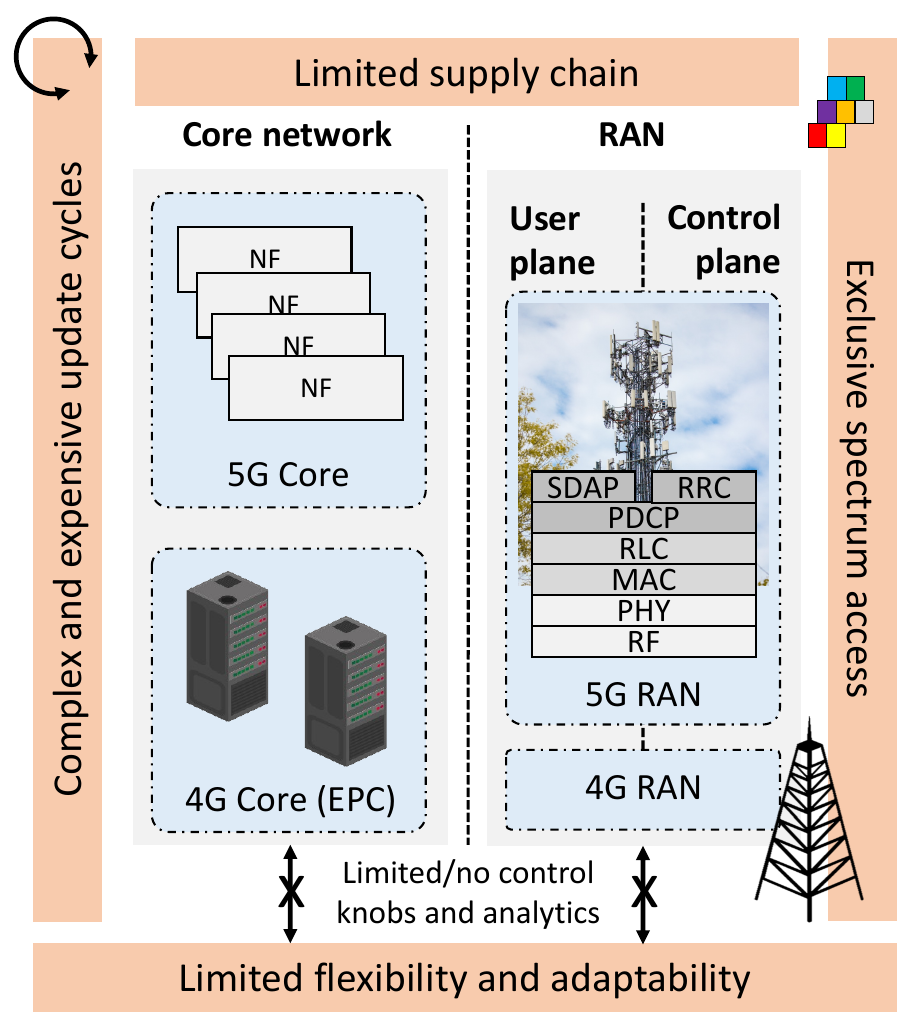}
    \caption{Components and abstractions of 4G and 5G cellular networks, with  the limitations associated to the current cellular architecture and deployment models.}
    \label{fig:limits}
\end{figure}

In this section, we review the high-level characteristics of cellular network architectures that have been typically deployed in 4G and 5G networks, primarily with \gls{3gpp} \gls{lte} and \gls{nr}~\cite{3gpp.38.300,parkvall2017nr}, and discuss opportunities for the ecosystem to accelerate the pace of innovation.

\subsection{Access and Core Network}

The first functional split in wireless networks is horizontal, i.e., two different parts of the system provide last-mile access to the users (i.e., the \gls{ran}) and support functionalities (i.e., the core network). 

\glsunset{cran}

As discussed in~\cite{parkvall2017nr,lien20175g}, the \gls{ran} comprises the set of base stations, either \glspl{gnb} for a 5G NR \gls{ran}, or \glspl{enb} for a 4G LTE RAN. The base stations provide wireless access to the mobile network to \glspl{ue}; in 4G, they are generally deployed as monolithic units, and in 5G as virtualized (vRAN) and even cloudified (C-RAN) instantiations. As shown in Fig.~\ref{fig:limits}, each base station is equipped with a complex protocol stack defined by the 3GPP, where the physical layer takes care of \gls{dsp}, channel estimation, transmission and reception, and the data link layer, in charge of scheduling and \gls{qos} optimization, is broken down into multiple sub-layers. The \gls{mac} layer handles scheduling; the \gls{rlc} layer buffering, concatenation, and segmentation; the \gls{pdcp} layer performs data encryption and packet sequencing; the \gls{sdap} layer enforces \gls{qos}; and the \gls{rrc} layer implements the state machine of the network, among other things~\cite{3gpp.38.300}. This stack has evolved across different generations of \gls{3gpp} specifications, e.g., the \gls{sdap} layer has been introduced in NR for \gls{5g} systems.

The core network primarily manages user authentication, mobility, paging, and routing to and from the public Internet. Fig.~\ref{fig:limits} reports a high-level illustration of the the \gls{4g} core, or Evolved Packet Core~\cite{3gpp.23.002}, and of the 5G core, which is based on chaining and orchestrating multiple atomic network functions through a \gls{sba}~\cite{condoluci2018softwarization}. Whilst the 4G core is typically deployed via monolithic servers, the 5G core is fully virtualized and can be deployed via virtual machines or Kubernetes-based micro services. 
The \gls{sba} approach, however, has not yet fully realized its potential.

\subsection{User and Control Plane}

\glsunset{up}
\glsunset{cp}
\glsunset{upf}
The second functional split is vertical, i.e., there are different procedures, protocols, layers, and core network components to manage the transmission and reception of user data (user plane) and the connectivity lifecycle (control plane). This separation introduces a clean distinction between configuration, optimization, and management, which are under the purview of the control plane, and the processing of the data in any shape (e.g., bits, packets), which is handled by the user plane.

The user plane processes data hierarchically. First, it manages IP packets, for example by applying encryption at the \gls{pdcp} layer and associating \gls{qos} levels to end-to-end packet streams, or bearers, at the \gls{sdap} layer. Then, packets buffered at the \gls{pdcp} and \gls{rlc} layers, which segment and concatenate them at a byte level into transport blocks, based on the resources scheduled by the \gls{mac}. The physical layer then transforms them into a waveform by processing the input at a bit level. These operations are performed in reverse order at the receiver. In the core network, specific functions (e.g., the User Plane Function, \gls{upf}, in the 5G core) act as packet gateways between the public Internet and the telecom operator network, and establish bearers with the mobile devices.

The control plane guides the components of the user plane to properly perform their operations, and manages user sessions and mobility. In the \gls{ran}, the control is orchestrated by the \gls{rrc} layer, which implements a finite state machine with well-defined transitions to describe the state evolution for connected users, from their initial access to connected mode and mobility events. The core network also has control plane elements that interact with the \gls{rrc} to authenticate the user, manage billing, and track its location in the network.

\subsection{Limitations and Theory-Implementation Gap}

These cellular network design principles are well-thought and the outcome of years of evolution of cellular systems, with 5G networks reaching hundreds of Mbps of throughput in typical scenarios~\cite{neutral2018,shafi2017deployment,narayanan2022comparative}. However, there is still a gap between cellular deployments and the rich set of features available in the specifications from 3GPP and in the technical literature from the wireless and networking research communities. The capabilities of cellular networks are today mostly focused on broadband applications, and lag behind the vision of what applications could achieve in mobile scenarios:

\begin{figure*}
    \centering
    \includegraphics[width=.85\textwidth]{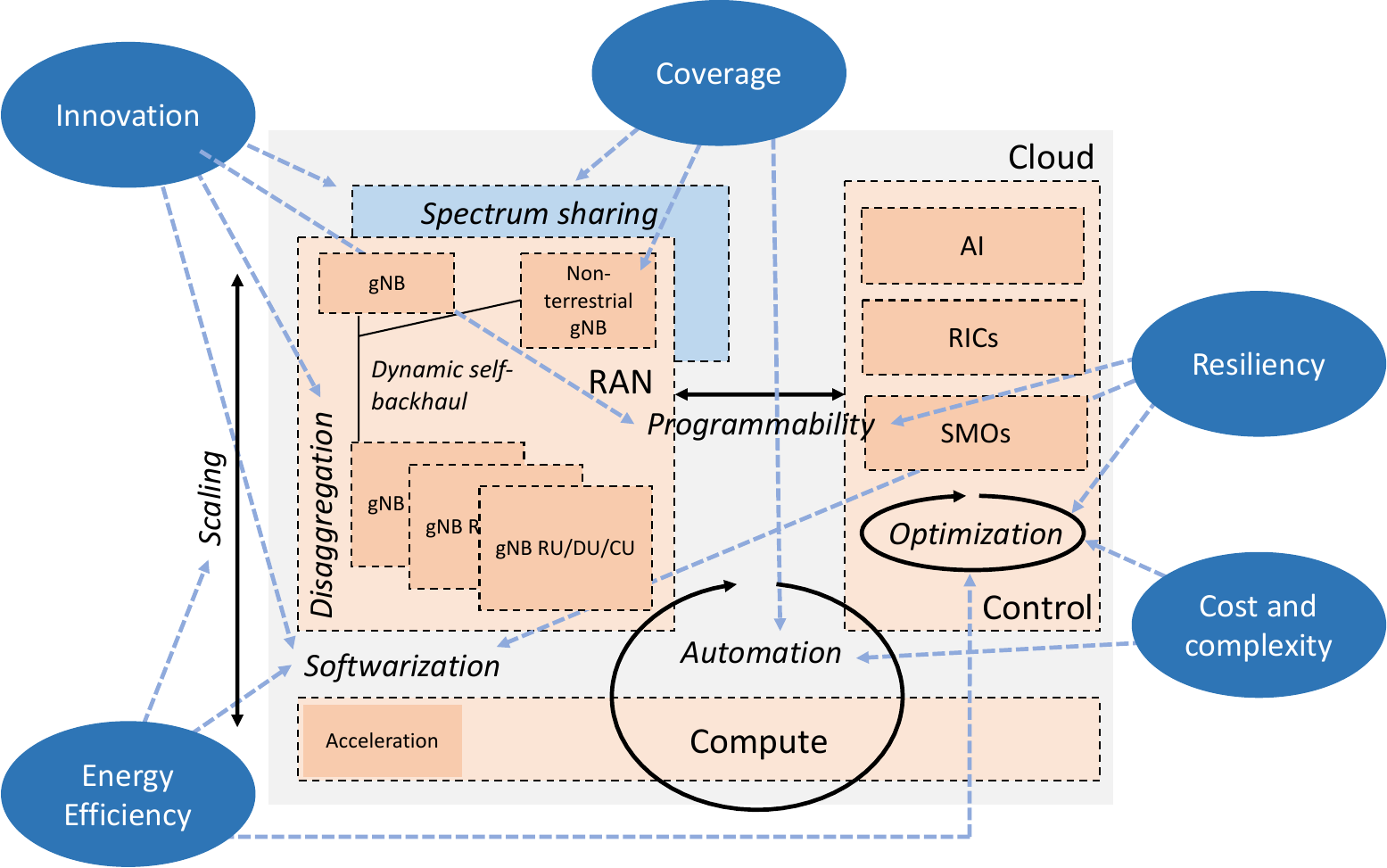}
    \caption{Foundational architectural principles of next-generation cellular networks (black), overlayed on the Open \gls{ran} architecture (orange), and with 6G target \glspl{kpi} (blue) related to energy efficiency, cost and complexity, innovation, coverage, and resiliency.}
    \label{fig:principles}
\end{figure*}

\begin{itemize}
\item First, the complexity, scale and performance requirements as well as lack of a clear monetizable use-case of production-grade cellular networks lead operators and thus vendors to focus on a subset of well-tested features to mainly support mobile broadband applications. This limits the potential of cellular network deployments, as well as the number of parameters that can be tuned and optimized in production. A further setback is that very few operators have upgraded to 5G standalone networks, complicating the support for different kinds of use cases at the same time, e.g., ultra-low latency data streaming for industrial control on a public network while also serving video streaming to residential users.

\item Second, the limited flexibility  extends to spectrum, the most valuable resource when deploying wireless systems. Cellular networks, so far, have been mostly deployed on licensed spectrum, which provides guarantees for no or limited interference, but also limits the bandwidth to small chunks and disaggregated chunks of spectrum (at least below 6 GHz). More flexible radios and spectrum access systems can pave the way for more efficient spectrum utilization and an overall increase in the available bandwidth.

\item Third, there are limited options for the deployment of the network itself. Whilst 3GPP interfaces and the standard itself are completely open, market consolidation has led to a small set of RAN vendors.
% Note that these numbers are similar to other critical technologies, like cloud.
Whilst vendor interoperability is ensured via well-tested Xn interfaces, further disaggregation could potentially spurn more innovation and also increase the resiliency of the supply chain~\cite{pinson20235g,ntiaNofo,germanSecurity}.  

\item Finally, whilst the 5G architecture allows for cloudified deployments, which could simplify continuous cycles of integration and deployment, the telco requirements are so stringent that the general availability of production-grade  cloud fabric has taken longer than anticipated. 
\end{itemize}

Overall, complexity, lack of adaptation, and limited flexibility offset the benefits introduced by the robust protocol stack design of cellular systems, and potentially prevent the adoption of state-of-the-art techniques to provide more innovative user services. This challenges the emergence of new market entrants for private and public cellular systems, limiting the diversity of the supply chain and telecom ecosystem and stymieing fast-paced competition and innovation.

\section{Empowering the 6G Cellular Architecture with Open \gls{ran} Principles}
\label{sec:new-arch}

In this section, we describe how 6G can benefit from the adoption of a new, system-level approach for the design and deployment of the network architecture, leading to a streamlined transition of innovation from research and standardization to production. It is important to distinguish between standards for key functionalities (as mainly done by the 3GPP); for enabling a softwarized/cloudified deployment with closed-loop control (e.g., O-RAN Alliance); and the actual deployments.
% (which could be SBA implemented via virtual machines or containers). 

The quest that led the development of 4G and 5G networks was driven by improvements in spectral efficiency (e.g., higher order modulations and new coding schemes), access to spectrum (e.g., new frequency bands at \glspl{mmwave}), and support for machine connectivity and low-latency systems. Future wireless systems will also target improvements in complementary \glspl{kpi}, as shown in the blue circles in Fig.~\ref{fig:principles}:

\begin{itemize}
    \item \textbf{Energy efficiency} --- energy consumption is one the main cost drivers for telecom operators (up to 60\% of the operational expenses~\cite{gsma2020energy}), and, overall, the information and computing technologies industry contributes to two to three percent of the total greenhouse gases emission~\cite{FREITAG2021100340}. To this end, recent literature has focused on improving the energy efficiency as one of the key elements of the design of next-generation cellular systems~\cite{hu2021energy,zhang2021guest,elsayed2019ai}, improving radio components, protocols, efficiency, and utilization.

    \item \textbf{Lower deployment and operational costs and complexity} --- upgrading cellular networks across generations is a complex and costly exercise for network operators, with cycles that take years to complete (e.g., one major U.S. operator moved commercial traffic to a 5G standalone core network in 2022, three years after launching a commercial 5G service~\cite{kirby20235g}). Similarly, operations require careful planning and supervision with humans in the loop, increasing expenses and limiting the scope of optimization to few selected parameters and services. Future wireless systems will need streamlined and automated processes for network deployment and operations.

    \item \textbf{Faster innovation cycles} --- once a network is operational, deploying and testing new features and functionalities, e.g., features from new 3GPP releases, represents an additional effort and cost, as it could lead to \gls{sla} requirements violation and risk of service disruption. This slows down innovation and prevents a quick adoption of new techniques and solutions in cellular networks---a key challenge to address in next-generation networks to unlock faster transition from lab research to cutting-edge commercial products.

    \item \textbf{Ubiquitous coverage} --- the deployment of high-frequency 5G networks  has focused so far on dense urban markets, as the anticipated utilization in suburban and rural deployments does not justify the deployment cost. This creates cellular networks evolving in two diverging tracks, with less densely populated areas left behind. In addition, significant portions of the world still lack cellular connectivity. Providing ubiquitous coverage is one of the targets of next-generation wireless systems, with research that so far has focused on non-terrestrial systems and reducing cost for deployment~\cite{giordani2021non,hokazono2022extreme,azari2022evolution}.

    \item \textbf{High resiliency} --- the cellular infrastructure is critical to our society, and major downtime of a carrier infrastructure often causes significant losses for various sectors of the economy~\cite{outages2020}. Therefore, increasing the resiliency of the network infrastructure to different incidents (either software-based, or caused, for example, by power outage or weather events) is paramount as we make cellular connectivity even more diffused and essential.
\end{itemize}

This is in addition to further improvements in throughput and latency, e.g., through ultra-wide band networks in the sub-terahertz spectrum~\cite{polese2020toward,akyildiz2022terahertz}. To achieve these target, there is an opportunity to rethink the cellular architecture and operations with a holistic approach which can provide more gains than the  improvements in the lower layers of the \gls{ran} protocol stack (e.g., the redesign the physical layer with \gls{ai}~\cite{oshea2017introduction,oshea2016learning}). 

\subsection{6G requirements and Open \gls{ran} principles}

Figure~\ref{fig:principles} illustrates how the \glspl{kpi} discussed above can be supported by foundational principles of Open \gls{ran} and next-generation cellular networks. The figure envisions a software-based disaggregated Open \gls{ran} system which is scalable and flexible, automated and optimized, and capable to dynamically access licensed, unlicensed, and shared spectrum for access and self-backhaul links. 

Specifically, through \emph{softwarization}, network functions are implemented as software and deployed on generic compute solutions, generally coupled with accelerators for digital signal processing. Via \emph{disaggregation}, monolithic network components are split into atomic network functions, enabling easier access to market at the cost of additional integration testing.
% This
% enables faster innovation, as it lowers the barrier to entry of new components in the \gls{ran} and new players in the telecom market. Meanwhile, 
% Integration and verification become important necessities - bearing a cost which currently outweighs the advantages of disaggregation and thus needs to be addressed by the innovation ecosystem. 
Softwarization and disaggregation can potentially improve energy efficiency when combined with dynamic \emph{scaling} of resources and aggregation of \gls{ran} components at a data-center scale. Softwarization also improves the resiliency of the access to the spectrum, using, e.g., agile software-defined radios \cite{dressler2022physical}, as well as of the network infrastructure. Indeed, as done in 5G today, software components implementing network functions can be deployed on generic hardware, which can be easily replaced, and quickly transferred in case a compute node fails (e.g., via micro-services~\cite{dressler2022v-edge}).

\emph{Automation} streamlines complex operations through a declarative process where tasks are defined in advance and executed at runtime. It cuts cost and complexity, reducing the need for manual supervision and control, and decreases the time it takes to apply configurations, restore services, and operate the network. The operator can then express high-level intents to guide the automation framework towards tailored deployments and configurations~\cite{doro2022orchestran}. Automation also improves resiliency, when combined with softwarization and programmability, and coverage, as it simplifies the deployment, operations, and management of network solutions in remote locations and at an extended scale. Security, however, remains an important issue because of the increased attack surface. 

\emph{Optimization} provides configurations for programmable \glspl{ran} based on detailed and realistic representations of the network state. This improves the resiliency of the network, which can self-optimize in case of failures or changes the operating scenario; the energy efficiency, as energy can be included as part of the target \glspl{kpi} to optimize; and the overall performance of the system, thus eventually reducing the cost per bit. \emph{Programmability} allows dynamic adaptation of \gls{ran}, compute, cloud, and backhaul networking functionalities through closed-loop control and standardized \glspl{api}. To this end, it fosters innovation and improves network resiliency enabling new control routines and adaptation strategies.

Through \emph{spectrum sharing}, next-generation cellular networks will provide a more flexible air interface and spectrum access mechanism, which can be designed in different ways (e.g., access to new bands~\cite{atis6gmidband}, unlicensed cellular~\cite{lagen:19}, sharing of cellular bands across different operators~\cite{tehrani2016licensed}, the neutral host model~\cite{LAHTEENMAKI2021102201}). This is associated with improved coverage, as spectrum can be allocated more efficiently based on demand and availability~\cite{bonati2023neutran}, and non-exclusive access to spectrum can create incentives for operators to deploy networks in remote or rural locations through reduced spectrum cost. In addition, non-exclusive access to the spectrum can enable new players and use cases, as for private 5G deployments, thus increasing competition in the cellular market.

With \emph{self-backhaul}, access and backhaul are multiplexed on the same waveform, protocol stack, and portion of the spectrum. It is another ingredient toward improving coverage, as it allows more flexible network topologies which are not constrained to the availability of fiber connectivity to the access nodes. Potentially, this can extend to network domains that are not traditionally considered in cellular systems, as \glspl{ntn}, thus further improving \emph{global} coverage. Wireless self-backhaul, in addition, can extend to additional portions of the network if compared to traditional inter-gNB or gNB-to-core backhaul, e.g., for interfaces across layers of the same gNB, gNB to edge deployments, or across different radio access technologies (e.g., Wi-Fi and cellular).

% \hl{Mention some non-obvious use cases that can be enabled by this architecture}
\textbf{How would an Open RAN 6G deployment look like?} Let us consider an example. In a rural location, two small operators deploy low-cost radios in the field, with small footprint and energy consumption. These are connected to a local edge data center through wireless backhaul, where base station components execute on generic compute resources. The two operators dynamically share the same portion of the spectrum, to improve coverage and availability of service to their users, and leverage slicing and other optimization to serve users with extremely heterogeneous requirements in mobile scenarios. The \glspl{ran} are connected to a remote core (running on a public cloud shared with other operators) through wireless backhaul. A centralized management and orchestration plane provides resiliency, optimization, and high service availability. The network self-adapts to traffic and usage patterns to minimize the energy consumption, e.g., adapting transmit power, number of network functions being executed, compute infrastructure actively used, and spectrum access. 

This, and other use cases with higher impact, are not well supported by the current cellular architectures, as discussed in Sec.~\ref{sec:trad-arch}, because of the reduced flexibility, automation, and optimization, calling for the integration of Open RAN principles in the architecture of next-generation 6G networks.
In the following sections, we discuss the foundational principles behind this architectural shift for cellular networks.

\section{Enablers: Softwarization, Programmability, and Virtualization}
\label{sec:program}

The transition to software-based, programmable, and virtualized environments is leading cellular networks into more flexible and dynamically tunable systems, which can benefit from fast deployment and reprogrammability cycles compared to traditional cellular deployment.

Softwarization is already mature in the wired Ethernet switching domain, where white-label switches can run different flavors of operating systems and programmable protocol stacks. 
% Whilst only a fraction of the market picked up on this innovation potential, t
This transition was spearheaded by seminal papers published since 2008, e.g., on OpenFlow~\cite{mckeown2008openflow}, which introduced the concept of \gls{sdn} and separation of control and switching for Ethernet campus networks; on operating systems for networks~\cite{gude2008nox}; and on programmable data planes based on the P4 language~\cite{bosshart2014p4}.

Cellular networks have been on the same evolution path since the introduction of the 5G Core, which, as discussed in~Sec.~\ref{sec:trad-arch}, adopts a softwarized service-based architecture.
% that disaggregates functionalities into atomic network functions that can be implemented via micro-services. 
This has opened the door to multiple open-source and software-only implementations of the core network itself~\cite{nextepc_website,open5gs_website,free5gc_website}, and softwarization is now being considered for the \gls{ran} of next-generation wireless systems. The community, however, has also realized that many of the core network functions only serve one or a very limited purpose, thus defying the need of an SBA. Clearly, more work is needed in 6G to ensure a proper way to disaggregate the functions.

The softwarization of the \gls{ran} has its roots in \gls{sdn}, on one side, and on \gls{cran}, on the other. The latter emerged as a paradigm to virtualize parts of the \gls{ran} computations in general-purpose data centers at the edge, with performance gains associated to centralized computing and control and energy efficiency due to dynamic scaling of the compute resources~\cite{checko2015cloud}. \gls{cran} has been widely studied in the literature and has influenced the roadmap for implementation of software-based cellular systems in current and next-generation cellular networks. The \gls{cran} Alliance, a group of operators pushing for the implementation and adoption of \gls{cran} systems, is one of the two entities that coalesced into the current O-RAN Alliance, together with the xRAN initiative~\cite{xran2018pr}.

\begin{figure}[t]
  \centering
  \includegraphics[width=\columnwidth]{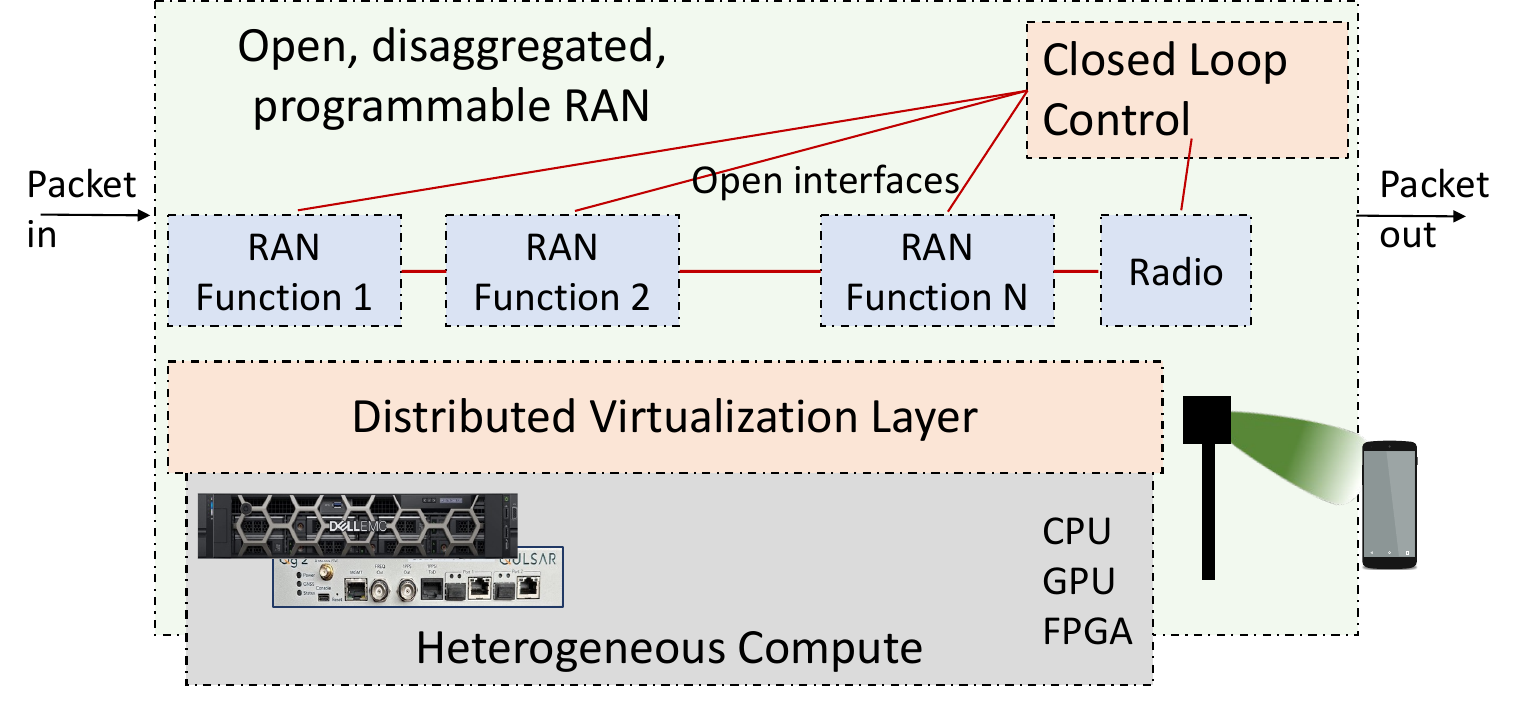}
  \setlength\belowcaptionskip{-.3cm}
  \caption{Representation of a cellular system where a virtualization layer enables software-based, programmable network functions.}
  \label{fig:virtualization}
\end{figure}

\begin{figure*}[t]
  \centering
  \includegraphics[width=.85\textwidth]{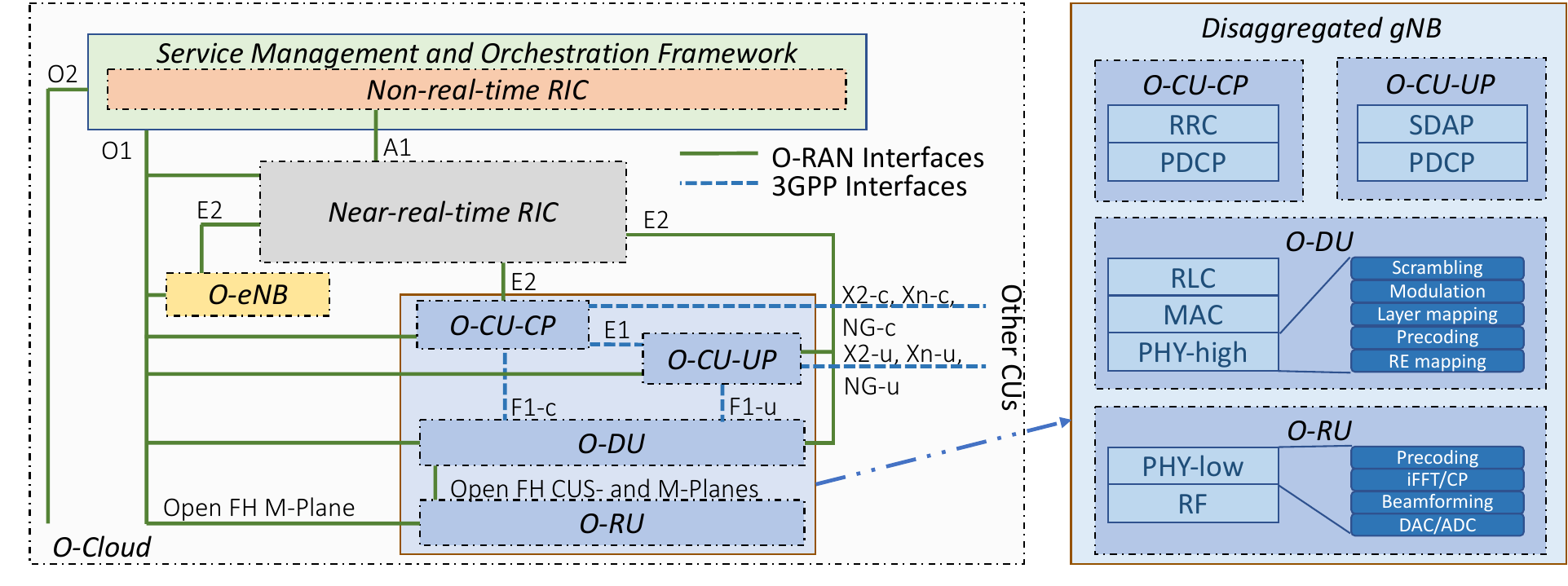}
  \setlength\belowcaptionskip{-.3cm}
  \caption{\gls{ran} disaggregation and open interfaces. Adapted from~\cite{polese2023understanding}.}
  \label{fig:disaggregated-architecture}
\end{figure*}

Through softwarization and virtualization, cellular networks can be deployed on white-box servers and radios, removing the tight coupling between software and hardware, as shown in Fig.~\ref{fig:virtualization}. 
This presents several advantages, including the programmability of networks, as it becomes easier to update configuration of  cellular networks which are not anymore bound to hardware parameters and constraints, but implemented as software-based network functions. 
Other advantages include the diversification of the supply chain ecosystem in wireless networks, as software components have a lower barrier of entry if compared to hardware implementations of a protocol stack. 
It also decreases the time from idea development to prototyping and implementation, as transitioning software across domains is a streamlined process if compared to developing hardware systems. 
Further, a software-based approach allows adopting best practices from the software industry, including cloud compute solutions, micro-services for resilient and easily scalable platforms, and fast deployment cycles with quality guaranteed by a continuous testing, integration and deployment cycle, as we discuss in Sec.~\ref{sec:cloud}.

On the other hand, softwarization, as discussed above, introduces performance challenges related to the implementation and performance tuning of \gls{dsp} routines. This has led to (i) effort to define minimum and desirable performance requirements for whitebox hardware that supports Open \gls{ran} deployments; and to (ii) the emergence of multiple \emph{programmable} solutions for \gls{dsp} acceleration, including \glspl{gpu}~\cite{kundu2023hardware}, \glspl{fpga}~\cite{karle2022hardware}, or dedicated systems-on-a-chip~\cite{edgeq}. Extending generic compute with programmable accelerators combines reconfigurability with fast and massively parallel signal processing, and leads to platforms and systems which are easier to upgrade, maintain, and source when compared to dedicated \glspl{asic}.

\subsection{Implementation and Research Directions}

\textbf{Hardware Abstractions.} To facilitate implementation, the O-RAN Alliance \glspl{wg} 6 and 7 are focused on developing a set of specifications around an \gls{aal} for the \gls{ran} and white-box hardware requirements for specific network functions~\cite{oran-wg6-aal-fec,oran-wg6-aal-gap}. \glspl{aal} are designed to establish standardized \glspl{api} that facilitate communication between dedicated hardware-based processors and the O-RAN softwarized infrastructure. This interaction covers various functions, such as channel coding/decoding and forward error correction~\cite{oran-wg6-aal-fec,oran-wg6-aal-gap}. However, managing and exposing \glspl{gpu}, \glspl{fpga}, and CPUs or system on chips with varying capabilities with a shared \gls{api} is challenging because of their heterogeneity. The authors of~\cite{kundu2023hardware} review the approach adopted by the O-RAN WG 6, which is based on abstracting and providing access to functionalities required by the 5G physical layer, but future research could look into alternatives which are not tightly coupled to specific physical layer features but can be extended and supported across multiple generations.

\textbf{Networking, Compute, and Energy Efficiency.} The second research and implementation challenge is associated with how to maintain high efficiency, in terms of computing power, energy, and cost, when comparing to traditional, monolithic devices, which are often highly optimized and accelerated by dedicated \glspl{asic}. Possible approaches can rely on statistical multiplexing and scaling, so that savings are achieved on average compared to traditional systems, and on improvements and optimization related to software design and hardware platforms targeting telecom and \gls{dsp} workloads.

\section{Open RAN Principle \#1: Disaggregation and Open Interfaces}
\label{sec:disaggregation}

The transition to software-based platforms, where components can be implemented as software (possibly via micro-services) has also ushered in more degrees of freedom in how different layers and functionalities of the protocol stack are mapped into atomic network functions. Traditionally, and as discussed in Sec.~\ref{sec:trad-arch}, 4G base stations are largely deployed as monolithic units, i.e., with all the functionalities possibly virtualized but in a single unit in proximity to the cell site. 

Figure~\ref{fig:disaggregated-architecture} illustrates how the softwarization of the 5G \gls{ran} helps organizing the \gls{gnb} functionalities into logical units, the \gls{cu}, \gls{du}, and \gls{ru}. The \gls{cu} is further split across the \gls{up} and \gls{cp}. The right part of Figure~\ref{fig:disaggregated-architecture} show how different protocols in the \gls{3gpp} stack are distributed across the units. The \gls{cu}-\gls{cp} maintains the state machine of the network, keeping track on the users state at the \gls{rrc} layer. The \gls{cu}-\gls{up} hosts the \gls{sdap} and \gls{pdcp} layers for managing data radio bearers. The \gls{du} features three layers which operate in a tightly synchronized fashion, and would thus be challenging to distribute across non-colocated network functions---these include \gls{rlc}, \gls{mac}, and the higher part of the physical layer. The latter is split based on operations that are carried out in the frequency domain (scrambling, modulation, mapping to \gls{mimo} layers, precoding, and mapping to resource elements), and in the time domain, which are implemented in the \gls{ru}. The latter also features the \gls{rf} domain components for signal digitalization, upconversion, and, in case, beamforming.

3GPP and O-RAN Alliance interfeces ensure that these components are connected together and to the O-RAN \glspl{ric}, as shown in the left part of Fig.~\ref{fig:disaggregated-architecture}. The \gls{cu}  and the \gls{du} are connected through the E1 and F1 interfaces, respectively, defined by \gls{3gpp}~\cite{3gpp.38.470,3gpp.38.300}. The \gls{cu} also supports connectivity (e.g., for handover) to other \glspl{cu}, \glspl{enb}, and to the core network components through the Xn, X2, and NG interfaces, respectively, also defined by the \gls{3gpp}. 

The \gls{du} is connected to one or multiple \glspl{ru} through the O-RAN Fronthaul interface, which builds on an enhanced version of the common public radio interface called eCPRI. It provides a reliable and synchronized transport layer for control messages and user data~\cite{oran-wg4-fronthaul-cus}. The fronthaul interface is implemented through four planes. The user plane carries in phase and quadrature (I/Q) samples corresponding to transport blocks to be transmitted. The control plane is used to signal when these I/Q samples need to be transmitted, and to anticipate uplink and random access slots in which the \gls{ru} needs to process incoming signals. The synchronization plane takes care of maintaining a nanosecond-level synchronization between the \gls{du} and the \gls{ru} through the \gls{ptp}. Finally, the management plane connects the \gls{ru} to the \gls{smo} (either directly, or through the \gls{du}), and can be used to push configurations to the radio, including, for example, beamforming codebooks~\cite{oran-wg4-fronthaul-m}.

Three other O-RAN interfaces include E2, O1, and O2. The first connects the \nearrt \gls{ric} to the \glspl{cu} and the \glspl{du}, to retrieve telemetry and \glspl{kpm} and to enforce control or apply policies in the \gls{ran} nodes~\cite{oran-wg3-e2-gap}. Data processing and control is performed by plug-and-play components within the \nearrt \gls{ric}, i.e., the xApps. The E2 interface is designed to be flexible and extensible, with an underlying application protocol (E2AP) to manage the connection lifecycle, and service models (E2SMs) developed on top to provide actual functionalities. For example, E2SM \gls{kpm} is used to stream telemetry, while the E2SM \gls{rc} and \gls{ccc} are used for control of \gls{ue}-specific and cell-specific parameters, respectively. 
The O1 interface connects the \nonrt \gls{ric} and the \gls{smo} to all the relevant \gls{ran} functions, as shown in Fig.~\ref{fig:disaggregated-architecture}. It is used to retrieve files, data, and configurations at a slower time scale compared to the E2 interface, and to push configurations and updates. It also maintains a heartbeat and provisions new services. The O2 is the interface between the \gls{smo} and the O-Cloud for service deployment on the cloud infrastructure.
Finally, the A1 interface connects the two \glspl{ric}, with the \nonrt \gls{ric} pushing policies and, if needed, external information elements to the \nearrt \gls{ric} and its xApps. 

\subsection{Implementation and Research Directions}

\textbf{Automated Interface Generation.} While openness allows for a mix and match of components and vendors, it comes with its own challenges associated to integration, interoperability, and interface generation. 
A key research direction involves understanding how \gls{ai} and natural language processing tools can be used to automatically generate the software implementation for the open interfaces based on their specifications.
%
% Therefore, a key research direction for Open \gls{ran} systems relates to how the openness of the interfaces can translate into automated processes for the generation of the interface implementation based on the standard specifications, for example by leveraging \gls{ai} and natural language processing tools.
%
Similarly, \gls{ai} can be used to automate testing and facilitate the integration of products from different vendors, identifying the incompatibilities, possible divergence of the implementations from the specifications, and security issues. Automated interface generation and testing decreases the time it takes from specification development to implementation availability, and reduces one of the pain and risk points associated to O-RAN, i.e., interoperability in a disaggregated environment.

\textbf{Extending E2 Service Models.} Similarly, automated interface generation can also encompass the E2 service models. As discussed above, the O-RAN Alliance has defined a basic set of service models for RAN control and streaming of telemetry and \glspl{kpm}. Additional service models, however, need to be developed to enable additional functionalities and control, as, for example, for spectrum sharing and \gls{iab} optimization (as discussed in Sec.~\ref{sec:iab}). 
A system design challenge involves creating service models for a wide range of functionalities and emerging use cases while being able to operate within parameters specified by the 3GPP for its \glspl{cu} and \glspl{du}. Additionally, it is important to define specific profiles and a fundamental set of features that must be incorporated by RAN equipment and software vendors to achieve O-RAN compliance.

\textbf{Interface Efficiency.} An additional research area is related to the efficiency of the data and control communications over the open interfaces, especially for those which are data-rate intensive and extremely sensitive to timing, e.g., the fronthaul interface. In this case, the data rate of the interface should not scale linearly with the bandwidth used for wireless communications, e.g., using advanced compression and aggregation techniques, to enable efficient massive \gls{mimo}, bandwidth scaling beyond the 400 MHz of 3GPP NR, or carrier aggregation. Theoretical models are also needed to understand the impact on latency and energy efficiency as compared to monolithic architectures, considering both functionalities in 3GPP user and control planes and control and optimization introduced by Open \gls{ran} elements.

% is an example of this, thus it is important to evolve its capabilities (e.g., when accounting for massive \gls{mimo}, bandwidth scaling beyond the 400 MHz of 3GPP NR, or carrier aggregation) without increasing the data rate over the interface linearly with the bandwidth used for the wireless communication, e.g., . 

\section{Open RAN Principle \#2: Toward Cloud-native Approaches for the \gls{ran}}
\label{sec:cloud}

Above-discussed softwarization and virtualization are positioning the \gls{ran} as a promising implementation candidate to reap the benefits of cloud-native software principles, automation, orchestration, and security. It is important to develop a shared and unified cloud abstraction that encompasses all the hardware components required to execute, optimize, and manage the \gls{ran}. Such cloud can span one or multiple locations over cell sites, the edge, regional data centers, and beyond~\cite{oran-wg1-arch-spec}. The O-RAN O-Cloud can support the deployment of all the components of an O-RAN system (e.g., those shown in Fig.~\ref{fig:disaggregated-architecture}), and embeds generic compute resources as well as accelerators for \gls{dsp} and hardware for AI/ML training, making it a hybrid and heterogeneous cloud platform~\cite{jain2013network} designed specifically for virtualization challenges associated to Open \gls{ran}. The O-Cloud does not only include hardware, but also software (e.g., hypervisors or container engines), the \gls{smo}, and its O2 interface~\cite{oran-wg6-o-cloud}.

Cloud-native principles transition cellular architectures into fully software driven solutions (except for the \gls{rf}-related components, e.g., antennas, \gls{rf} chains, and data converters). This comes with multiple benefits, including the possibility of automating the provisioning and management of \gls{ran} functionalities; multi-tenant \gls{ran} environments and neutral host solutions (e.g., as discussed in Sec.~\ref{sec:sharing} and~\cite{bonati2023neutran}); and the definition of a shared abstraction over different classes of heterogeneous hardware.

\vspace{-.15cm}
\subsection{Orchestration and Automation}
\label{sec:orchestration}

The provisioning of open interfaces and the introduction of a unified abstraction for the hardware open new opportunities for orchestration and automation of the whole \gls{ran} through \gls{ci}, \gls{cd}, and \gls{ct}. \gls{ci} ensures that new software patches or features are automatically integrated with the rest of the cellular network codebase. \gls{cd} automates the deployment of these features on the cellular network infrastructure. Finally, \gls{ct} continuously evaluates the performance, security, and compliance of the software, without the need for manual tests which may miss important changes or bugs.

\gls{ci}, \gls{cd}, and \gls{ct} are best practices adopted in cloud-native environments, enabled by a variety of workflows such as GitOps. GitOps is a state-of-the-art methodology for orchestrating \gls{ci} and \gls{cd} and efficiently managing infrastructure, which can be applied in O-RAN virtualized and cloud-based environments, as  shown in~\cite{bonati2023neutran}. At its core, GitOps relies on \texttt{git} repositories to serve as the authoritative source of truth, not only for the cellular network code, but also for the infrastructure configurations (e.g., \gls{ru} parameters, O-Cloud settings, among others). 
This approach emphasizes the use of declarative configuration, where the desired state of the system is clearly defined, making it easier to understand, review, and audit events and configurations of the cellular network.
Automated synchronization tools or controllers (e.g., ArgoCD) continually monitor these \texttt{git} repositories for changes and automatically apply them to the target environments, ensuring that the actual system state aligns with the defined state in \texttt{git}. 
By tracking infrastructure configurations and only allowing automated updates from authoritative sources, GitOps enables rapid and reliable application delivery while minimizing the risk of configuration drift. This methodology also fosters collaboration, code review, and security practices, as changes to infrastructure and configurations undergo the same scrutiny as code changes. 
Additionally, Git\-Ops supports observability and monitoring to uphold performance and reliability standards, facilitates the management of multiple environments (e.g., as in a distributed system such as the O-Cloud), and promotes portability across different hardware implementations. 
% making it an invaluable approach in cloud-native and containerized environments for efficiently managing complex, distributed cellular network.

\vspace{-.15cm}
\subsection{Security}
\label{sec:security}

Open RAN security is at the forefront of the discussion around open cellular systems, as the new network interfaces, the virtualization, softwarization, and usage of AI/ML for control can extend the threat surface of cellular networks. 

In terms of implementation challenges, the O-RAN Alliance \gls{wg} 11 has developed a comprehensive set of specifications that analyze the stakeholders and threat models for Open \gls{ran} systems~\cite{oran-sfg-requirements,oran-sfg-protocols,oran-sfg-threat}. 
These documents highlight how Open \gls{ran} has expanded the range of stakeholders responsible for the security of the \gls{ran} well beyond the confines of conventional 4G and 5G networks. Beyond traditional players like vendors, operators, and system integrators, accountability is now also required for network functions and virtualization platform providers, third-party developers, O-Cloud service providers, and administrators overseeing virtualized components. 

As for the threat surface, documents~\cite{oran-sfg-threat} and~\cite{eu2022cybersecurity} from the O-RAN Alliance and the European Commission have identified seven distinct threat categories, encompassing a wide array of attacks targeting different facets of the network. These threats span from attacks on the O-RAN infrastructure itself (e.g., the \glspl{ric}), to vulnerabilities affecting the O-Cloud, open-source code, physical infrastructure, wireless functionalities, protocol stack, and AI/ML components. Attacks listed in the literature can compromise the availability, integrity, and confidentiality of the network and its data~\cite{oran-sfg-threat, eu2022cybersecurity}. These attacks are associated with critical assets related to interfaces, data, and logical components, but also subpar product quality, underdeveloped technical specifications, supply chain tampering, and infrastructure failures~\cite{altiostar2021wp,ericsson2021securitywp,mimran2022evaluating,eu2022cybersecurity}.

The inherent openness of the platform, however, also enables operators to deploy tools and audit components for security---a task that proves challenging in closed solutions by vendors. The \gls{smo} assumes a crucial role in fortifying network operations, as it comes with a global view of the RAN network infrastructure and performance, and can run routines to spot anomalies and weaknesses in the system~\cite{altiostar2021wp}. 
Similarly, the automation and CI/CD/CT discussed above enable continuous testing and updates of the software, seamlessly deploying security patches in a timely fashion. Finally, technical specifications have also been issued to enforce authentication and encryption procedures, directly addressing security concerns unique to the O-RAN architecture~\cite{oran-sfg-protocols, oran-sfg-requirements}. As discussed in~\cite{groen2023implementing},  encryption impacts the performance of the O-RAN interfaces, showing that it is not a concern for the non-data-intensive E2.

\subsection{Deployment and Research Directions}

\textbf{Cloud-Native Approaches to Energy Efficiency.}
Energy efficiency is a key trend when it comes to cloud-native future network deployments. The virtualization of RAN components---combined with intelligent automated orchestration---enables the dynamic adjustment of compute resources to meet user requirements, thereby limiting power consumption to the specific network functions in use. This concept has been discussed in previous studies~\cite{sabella2014energy,garciaaviles2021nuberu}. Moreover, the closed-loop control capabilities mentioned earlier, combined with RAN virtualization, facilitate more precise and flexible sleep cycles for base stations and RF components. These components typically account for the majority of power consumption in cellular networks, as highlighted in various surveys~\cite{pamuklu2021reinforcement, luo2018reducing, lopez2022survey}. Future research and development efforts should focus on bridging the gap between state-of-the-art literature on energy efficiency and approaches that leverage and can be deployed on the open and virtualized components of Open RAN systems, going beyond what was possible in yesterday's monolithic, non-programmable deployments.

\begin{figure*}[t]
    \centering
    \includegraphics[width=.95\textwidth]{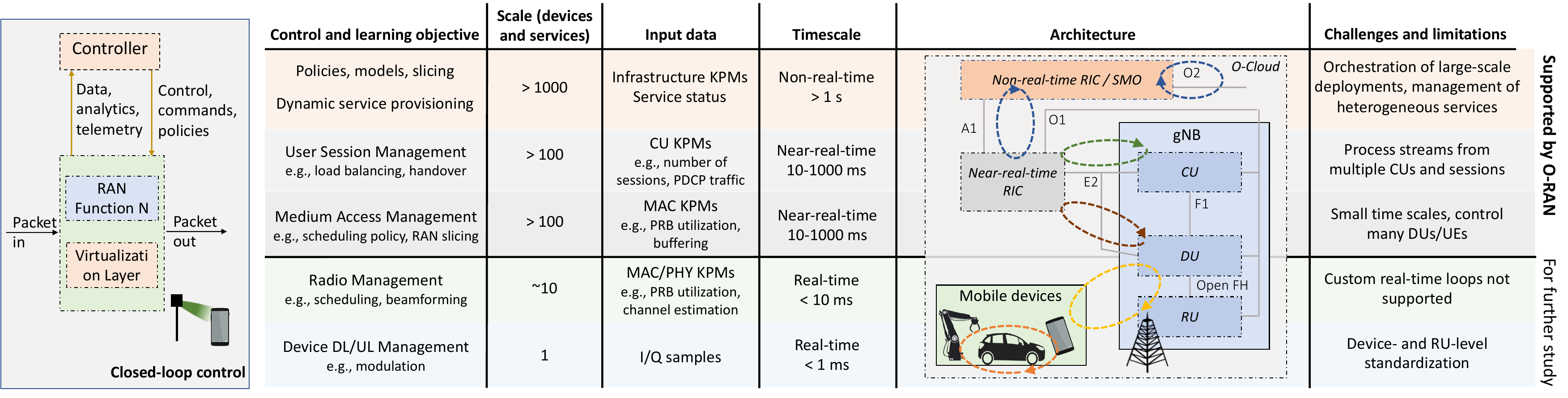}
    \caption{Closed-loop control in the Open RAN architecture. Based on~\cite{polese2023understanding,bonati2021intelligence}.}
    \label{fig:closed-loop}
\end{figure*}

\textbf{An O-Cloud with Heterogeneous Compute and Devices.} Similarly to the need for supporting and abstracting different hardware accelerators for virtualized \gls{ran} \gls{dsp}, discussed in Sec.~\ref{sec:program}, there are also challenges when it comes to defining an O-Cloud spanning heterogeneous hardware platforms and components. In this sense, the O-RAN Alliance \gls{wg} 7 is outlining the specific criteria that must be met by white box hardware to facilitate the implementation of O-RAN-compliant equipment, specifically when considering devices with \gls{rf} capabilities. This equipment encompasses various types such as indoor picocells, outdoor microcells, and macrocells, all operating within sub-6 GHz and mmWave frequency ranges. Additionally, it includes integrated access and backhaul nodes, as well as fronthaul gateways, all part of the architectural components depicted in Fig.~\ref{fig:disaggregated-architecture}. The specifications provide clarity on functional parameters relevant to specific use cases (e.g., frequency bands, bandwidth, inter-site distance, \gls{mimo} configurations) and outline the hardware attributes (e.g., accelerators, computing capabilities, connectivity) of these nodes.

\textbf{Automated Security for the RAN.} While the Open \gls{ran} paradigm enable visibility and security best principles, it is necessary to design, develop, deploy, and test the algorithms and infrastructure to actually do this. A promising research direction focus on understanding the role that \gls{ai} and \gls{ml} play to secure closed-loop control and the overall Open RAN infrastructure, e.g., through automated anomaly detection. The literature in the last decade has mostly focused on anomaly detection for wired networks~\cite{bhuyan2014network}, where \gls{sdn} and packet switching architectures provide enough insights to identify and flag anomalous traffic flows and react with access control mechanisms. The openness and interfaces of Open RAN systems can be leveraged to extend automated anomaly detection in the cellular domain~\cite{sundqvist2023robust}, thanks to the privileged point of view that the \gls{smo} and the \glspl{ric} have on the cellular infrastructure.

\textbf{Security for the AI/ML Control.} AI and ML themselves, however, are also open to vulnerabilities, which can manifest before and after deployment. Before deployment, tampering with training data can affect the training process itself. Similarly, backdoors can be added to deep learning models to manipulate model responses to specific inputs. After deployment, vulnerabilities can arise through adversarial machine learning and by corrupting input data during inference. Therefore, it is necessary to research resilient AI/ML solutions for Open RAN that safeguard against vulnerabilities stemming from data manipulation, as, for example, employing techniques such as autoencoders (as demonstrated in~\cite{groen2023implementing}), contrastive learning~\cite{kim2020adversarial}, and contaminated best arm identification~\cite{altschuler2019best}, among other techniques.

\section{Open RAN Principle \#3: AI-based Closed-loop Control} 
\label{sec:closed-loop}

The programmability of the network stack, the open interfaces, and the introduction of the \glspl{ric} are crucial component for the main transformation brought along by Open RAN: data-driven closed-loop control and optimization of the \gls{ran} functionalities. This is a fundamental step toward enabling autonomous networks that can self-adapt to dynamic requirements and deployment conditions. Whilst 3GPP offers mechanisms for closed-loop control, the O-RAN Alliance was the first to introduce a formal framework through the near-real-time and non-real-time RICs and associated interfaces. 

Figure~\ref{fig:closed-loop} illustrates how closed-loop control is enabled by, and integrates into the O-RAN architecture. A specific \gls{ran} function (e.g., the \gls{du}) exposes telemetry, analytics, and \glspl{kpm} through the O-RAN interfaces. These are received by a controller, which performs specific tasks based on the data it received from the network function. This task can be some classification or regression, to infer information on the \gls{ran} status; prediction, to anticipate dynamics in the network function (or in general, in the \gls{ran}) behavior; or control, leveraging reinforcement learning, optimization, or other data-driven techniques to determine the configuration of the network function that best matches the current status in the \gls{ran}. For the latter, control is sent back to the \gls{ran} function, thus actually closing the loop. 

This paradigm applies to different functions and optimization loops in the \gls{ran}, as shown in the right part of Fig.~\ref{fig:closed-loop}, which extends the analysis in our prior work~\cite{bonati2021intelligence,polese2023understanding}.
The controller can be a piece of custom logic executing in different components of the O-RAN architecture. Specifically, the O-RAN Alliance has drafted specifications for two \glspl{ric}, as discussed in Section~\ref{sec:disaggregation}, which cover different optimization domains, as shown in Fig.~\ref{fig:closed-loop}. The \nearrt \ric embeds custom logic through \emph{xApps}, while the \nonrt \ric uses \emph{rApps}. A third class of custom control logic is represented by \emph{dApps}, i.e., an extension of the O-RAN architecture proposed in~\cite{doro2022dapps} which represents custom plugins to be deployed alongside \glspl{cu} and \glspl{du}. 
The controller discussed above can be represented by any of the these applications (or a set of applications).

Different controllers focus on different control domains, based on their capabilities, access to data and \glspl{kpm}, and overall point of view on the network, as well as the timescale of the decisions. The combination of \gls{smo} and \nonrt \ric, together with their rApps, has a global point of view on the O-Cloud and on the services provisioned on the network. These components primarily focus on determining high-level policies and managing the lifecycle of network services with a non-real-time granularity, i.e., the control loop is closed after more than 1 s. While there is generally one \gls{smo} instance per network deployment, there can be multiple \nearrt \rics (with related xApps), which are deployed at the edge of the network and have visibility into a cluster of tens of \gls{ran} nodes. The control loops from the \nearrt \ric operate over the E2 interface at a timescale between 10 ms and 1 s, and influence the radio resource management process in the \ran with policies but also through the dynamic reconfiguration of \ran parameters (e.g., with E2SM \gls{rc} and \gls{ccc} as discussed above). Finally, dApps interface with a single \ran function at any given time, but are envisioned to be capable of performing control at a timescale below 10 ms (currently not considered by the O-RAN Alliance specifications).    

\textbf{Closed-loop Control Use Cases.} The flexibility and capabilities provided by the multiple control loops has led to research and development of rApps, xApps, and dApps for multiple use cases, toward the optimal configuration of O-RAN networks~\cite{polese2023understanding}.

Closed-loop control with Open RAN primitives allows for the fine-tuning of mobility management and performance for \emph{specific} mobile users, e.g., by adjusting handover, load balancing, multi-connectivity, access barring, and beamforming parameters within the \ran~\cite{oran-wg3-e2-sm-rc}. In~\cite{lacava2023programmable}, the authors show that tuning handover parameters with a data-driven loop that accounts for the bespoke requirements of individual \glspl{ue} improves throughput and spectral efficiency by an average of 50\% over traditional cell-based handover heuristics. This flexibility facilitates the optimization of the mobile experience for single UEs, opening new use cases and possible revenue streams for network operators.

Resource allocation is another crucial area where closed-loop control with Open \ran can outperform traditional approaches based on the point of view of individual \glspl{gnb}. Open RAN controllers can leverage the data and telemetry to understand user requirements and evolving contexts, and map them into effective configurations of the slicing and scheduling policies of the network which improve resource utilization and quality of service for users~\cite{zhang2022federated}. Researchers have explored the application of AI/ML-based optimization in network slicing, scheduling, and service provisioning, adapting the network to different slices and user needs~\cite{sharara2022cloud}.

This is an area where AI and ML have been widely used to drive the optimization. Experiments and demonstrations on experimental platforms like Colosseum and Arena testbeds~\cite{bonati2021colosseum,bertizzolo2020arena} have showcased xApps' capabilities to intelligently control the scheduling policies of various network slices on base stations~\cite{bonati2021intelligence,puligheddu2023semoran}. Different slices with specific optimization targets, such as \gls{embb}, \gls{urllc}, and \gls{mtc}, can be efficiently managed through closed-loop control mechanisms~\cite{polese2021coloran,johnson2021open,puligheddu2023semoran}.

The versatility of Open RAN extends to supporting new applications, such as vehicular communications and industrial \gls{iot} scenarios~\cite{abedin2022elastic,hoffmann2023signaling,linsalata2023oran,godor2020look}. Open RAN's capabilities, like dynamic control and adaptability of Massive MIMO configurations, can enhance mobile reliability and robustness~\cite{demir2023cell,linsalata2023oran,polese2018machine}. For industrial IoT applications, where high reliability and precise timing are crucial, Open RAN's closed-loop control can adapt configurations to the evolving conditions on factory floors~\cite{sathya2022toward,tselikis2023automated}.

Additionally, closed-loop control facilitates the optimization of the RAN deployment itself~\cite{doro2022orchestran,gramaglia2022netowkr,bega2020ai}. Researchers have proposed zero-touch orchestration frameworks, fault-tolerant techniques, and efficient matching schemes between different RAN components, all contributing to better resource utilization and overall network efficiency~\cite{tamim2021downtime,doro2022orchestran,li2021rlops,huff2021rft}.
Finally, security of the \ran can also be enhanced through closed-loop monitoring and control, as we discuss in Sec.~\ref{sec:security}. 

\subsection{Implementation and Research Directions}

\textbf{Towards A Single RAN.} The bulk of the 6G RAN and architecture standardization will be carried in 3GPP. However, whilst the O-RAN Alliance is mainly concerned with implementation and operations of networks, promising principles ought to be natively embedded into 3GPP standards efforts. More research is needed which maps efforts and roadmaps such that functional standardization (3GPP) can be aligned with operational capabilities (O-RAN Alliance).

\textbf{Expansion to New Use Cases.}
The current use cases, discussed above and in~\cite{oran-wg1-use-cases,oran-wg1-use-cases-analysis,polese2023understanding}, span various areas of radio resource management for cellular networks. However, as the capabilities of 3GPP RAN continue to evolve, encompassing scenarios like non-terrestrial networks and support for augmented reality/virtual reality (AR/VR) within the metaverse~\cite{chen2018virtual}, there arises a need for a more thorough refinement and evaluation of future use cases, considering the role that intelligent, data-driven closed-loop control can play in next-generation wireless applications.

\textbf{Efficient and Explainable AI/ML for RAN Control.}
While O-RAN provides the basic primitives to enable closed-loop control, how to achieve this in an intelligent and efficient way is an open challenge. The use cases discussed above rely on a mix of heuristics and AI/ML-based solutions. As industry transitions toward intelligent control, there is a need to identify robust, reliable, and deployable \gls{ml} solutions for wireless. In~\cite{polese2023understanding}, we discuss the \gls{ai}/\gls{ml} workflow that O-RAN systems support, covering the end-to-end process, from data collection to online inference and tuning. In this context, there are still open questions related to (i) optimal methods for training with offline data on systems which are fundamentally dynamic and online, but also very sensitive to performance degradation; (ii) \gls{drl} solutions that generalize well across deployments in different areas and with varying traffic distributions, and how generative \gls{ai} can be used in the Open RAN context; and (iii) explainability solutions for systems with dynamic control and complex input/output relationships, among others.

\textbf{Hierarchical Control.}
The possibilities led by two RICs and the dApps, and their capability to extract data from the network operations and to manage control decisions in the RAN, position them ideally for stemming AI/ML use cases. As mentioned above, AI/ML can optimize slicing decisions. Furthermore, recent studies show traffic steering and beamforming can benefit from RICs~\cite{dantas2023beam,lacava2023programmable} and mutually improve learning in the hierarchical structure of the RICs by the use of a novel ML technique called \gls{hrl}~\cite{habib2023hierarchical}. Different timescales of RICs make it challenging to have control-loops that use information from multiple timescales. However, certain network optimizations require fine \emph{and} coarse granularity data at the same time. HRL allows coordination of multiple timescales, yet, further research is needed to enable information fusion  and control in a smoother way than it is today.

\textbf{Conflict Mitigation.} This is a critical component when considering closed-loop control in Open RAN, especially in a hierarchical context as discussed above or with multiple xApps or rApps targeting the same base station. Indeed, there are instances in which different xApps may try to control the same parameter (direct conflict), or different parameters which have a correlated impact on the \gls{ran} (indirect conflict, e.g., one xApp reduces the resources associated to a slice while another xApp hands over multiple \glspl{ue} to that slice). Future research needs to evaluate and compare different conflict mitigation strategies, e.g., pre-deployment or post-deployment; based on explicit declaration of control policies or implicit reconstruction of the conflict impact or both; among other things. In addition, there is an open discussion in terms of which is the architectural component that is best positioned to perform conflict mitigation, e.g., the \gls{smo}, as discussed in~\cite{doro2022orchestran}, or the \nearrt \gls{ric}, or both.

\section{Emerging System Requirements \#1: Agile Spectrum and Infrastructure Sharing}
\label{sec:sharing}

Let us now explore two  system requirements which have been emerging recently as part of 6G design discussions. The first one, discussed in this section, is on spectrum and infrastructure sharing.
The demand for faster data rates and reduced latency in cellular networks has led to a significant increase in network densification~\cite{bhushan2014network}. This has also given rise to new deployment strategies, including wireless self-backhaul solutions (which we discuss in Sec.~\ref{sec:iab}), and to an increase in the number of private operators establishing dedicated cellular infrastructure~\cite{wen2022private}. Consequently, there is a significant portion of both capital and operational expenses faced by both public and private operators that goes towards acquiring access to spectrum, cell site facilities (like poles and towers), and equipment, as reported by the FCC~\cite{fccReport} and~\cite{oughton2018cost}.

These increased costs can be offset through spectrum and infrastructure sharing. The first has been recognized as an efficient means to enhance overall spectral utilization~\cite{zhang2017survey} and recent estimates indicate that the adoption of infrastructure and spectrum sharing techniques could potentially yield network operational cost savings of at least 30\% in the next five years~\cite{analysismason}. Infrastructure sharing is based on the neutral host model, where infrastructure is provided by third-party companies, leasing physical resources to multiple operators on a shared-tenant basis~\cite{LAHTEENMAKI2021102201,alpha2021}, reducing the overall infrastructure costs~\cite{samdanis2016network}.

As  discussed in~\cite{bonati2023neutran}, however, \ran and spectrum sharing are not yet suitable for widespread adoption in multi-operator network deployments~\cite{stl2022open}. This is primarily due to the absence of mechanisms that facilitate: (i) fine-grained sharing, where multiple tenants can share compute and spectrum slices from the same physical infrastructure; and (ii) dynamic sharing in licensed, unlicensed, or  partially licensed bands, which allows infrastructure owners to fully harness the statistical multiplexing of \ran and spectrum resources and adjust infrastructure parameters to meet tenant requirements that can change within seconds.
Consider, for example, spectrum sharing in the \gls{cbrs} band. This is a partially licensed band in the U.S. with different tiers of prioritized or general access to 150 MHz of spectrum within 3.55 GHz and 3.7 GHz, coordinated by a spectrum access system which currently operates on timescales in the order of minutes~\cite{palola2017field}. This limitation reduces system flexibility and eventually the efficiency in terms of spectrum utilization.

The openness and programmability introduced by the Open \ran paradigm have the potential to  upend how spectrum and infrastructure sharing are managed in practical deployments. Resource utilization is improved thanks to dynamic sharing, to the end benefit of the users which will be able to access more spectrum when needed. Specifically, virtualization and programmability principles can pave the way for automated and virtualized pipelines for the management of shared resources, offering a zero-touch, resilient, and fault-tolerant automation~\cite{bonati2023neutran}.

\begin{figure*}[t]
  \centering
  \includegraphics[width=.85\textwidth]{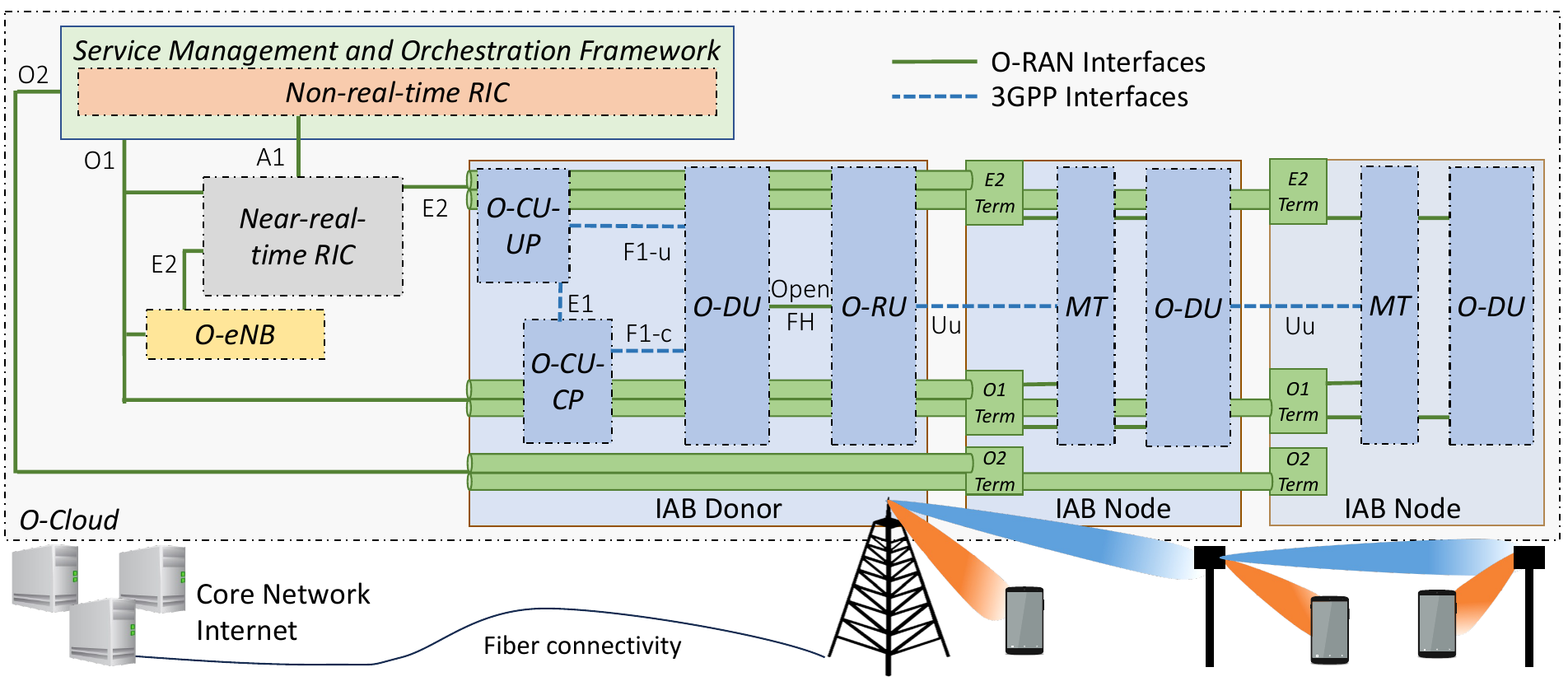}
  \setlength\belowcaptionskip{-.15cm}
  \caption{Proposed extension of the O-RAN architecture to managed self-backhaul scenarios based on \gls{iab}. Adapted from~\cite{moro2023iab}.}
  \label{fig:iab}
\end{figure*}

These functionalities play a crucial role in ensuring the reliability and effective coordination among multiple tenants, which can thus dynamically share infrastructure and spectrum resources without the need for manual intervention or over-provisioning, otherwise required in traditional cellular systems~\cite{garciaaviles2021nuberu,sarakis2021cost}. Additionally, they enable the timely and dynamic management of the lifecycle of network services. This has been a challenge in traditional cellular systems, especially when dealing with complex software services like softwarized \glspl{gnb} that need to be instantiated in a matter of seconds~\cite{liyanage2022survey,benzaid2020ai,doro2022orchestran}.

Note that the  O-RAN interfaces allow for better  accountability and visibility through fine-grained control over spectrum sharing~\cite{gsma2021sharing}. The Open RAN infrastructure also supports \gls{sla} enforcement through dynamic, fine-grained resource allocation, driven by optimization engines~\cite{sharma2017dynamic,bonati2023neutran,doro2022orchestran,foukas2019iris,caballero2019network,caballero2017multi}. This combination can enhance operators' confidence in shared systems and open up new possibilities for private and shared deployment scenarios.

\subsection{Research Directions}

Open-RAN-driven spectrum and infrastructure sharing can go beyond the capabilities discussed in the cognitive radio literature~\cite{liang2011cognitive,masonta2013spectrum} by combining data- and ML-driven approaches, a hierarchical control structure with decentralized and centralized endpoints, and automation and softwarization, with the end goal of developing sharing approaches that can interact with and are practically deployable in 3GPP cellular systems. Nonetheless, lessons learned and algorithms developed for cognitive radio systems can be analyzed and considered within the Open RAN framework. In addition, we identify two promising research directions as follows.

\textbf{Native Sensing and Sharing.} While Open \gls{ran} principles enable spectrum and infrastructure sharing, how to actually (i) sense and detect spectrum usage and allocations and (ii) distribute resources to the various tenants are still a challenge. Future research efforts can explore the practical implementation of Open-RAN-driven sensing, along with both reactive and proactive spectrum adaptation solutions, taking into account both 3GPP and non-3GPP systems, and licensed and unlicensed spectrum users~\cite{baldesi2022charm}. Additionally, it is essential to investigate extensions to the O-RAN architecture that are pertinent to spectrum sharing, e.g., dApps for real-time spectrum sensing~\cite{doro2022dapps}. 

\textbf{Spectrum Sharing in FR-3.} Another area of significant interest connecting 6G and intelligent spectrum sharing through Open \gls{ran} is sharing in the centimeter band, or FR-3, as identified by the 3GPP and recent literature~\cite{atis6gspectrum}. The frequency range between 6 and 24 GHz is of significant interest for 6G networks, as it presents a more favorable propagation environment if compared to the lower mmWave band in FR-2, and, at the same time, it can potentially accommodate wider channels compared to the overly crowded sub-6 GHz range. This spectrum, however, is currently allocated for key services for the military, weather and Earth monitoring, and satellite uplinks and downlinks, calling for a dynamic spectrum sharing approach to instantiate cellular systems when possible and avoid harmful interference to current incumbent.

\section{Emerging System Requirements \#2: Self-reconfigurable Wireless Backhauling}
\label{sec:iab}

As discussed in Sec.~\ref{sec:new-arch}, self-backhaul through wireless links is a key component of next-generation wireless networks, also considering the expansion to the non-terrestrial domain and the increasing densification of terrestrial access points in different frequency bands, including centimeter band (or FR-3) and millimeter waves (or FR-2). Since Release 16,  \gls{3gpp} has incorporated \gls{iab} into its specifications for NR~\cite{takao2020}. \gls{iab} supports the multiplexing of backhaul traffic with \gls{ue} access traffic over the same 5G NR air interface, forming a mesh of base stations connected wirelessly, without the need for expensive wired fiber optical connections to every access point, as shown in the bottom part of Fig.~\ref{fig:iab}. The \glspl{gnb} with a wired connection are the IAB-Donors, while the wireless relays are the IAB-Nodes~\cite{polese2020iab}. 

Compared to prior wireless mesh networks and relaying research, \gls{iab} brings opportunities related to being fully embedded within the 3GPP stack, including the possibility of using the same waveform and spectrum for access and backhaul, but also challenges for scheduling and providing reliability across both kinds of links. In this sense, despite reaching a mature standardization stage, there are ongoing challenges in managing, provisioning, and optimizing integrating access and backhaul. \gls{iab} provides optimization opportunities across all the layers of the 3GPP stack. At lower layers, specialized \gls{iab}-aware scheduling techniques are necessary to ensure fair and efficient resource allocation among UEs (in the access) and \glspl{mt} for IAB-Nodes, as discussed in~\cite{zhang2020sched,zhang2021resource}. Simultaneously, proper management of temporal and spatial scheduling for \glspl{iab} flows is vital to minimize interference~\cite{yu2023coordinated}. Furthermore, adaptive topology reconfiguration mechanisms are required to maintain resilience against link failures, traffic imbalances, and irregular user distribution, as explored in~\cite{ranjan2021cellselection}. These advanced management procedures demand control primitives beyond what the \gls{3gpp} has specified.

To this end, the Open RAN paradigm can introduce a shift in how \gls{iab} systems are managed, through programmability, softwarization, and disaggregation. The programmatic control of \gls{ran} components via open interfaces and centralized control loops, as described in~\cite{polese2023understanding}, holds significant potential for optimizing and managing \gls{iab}. The authors of~\cite{moro2023iab} discuss how the existing O-RAN architecture can evolve to accommodate \gls{iab} control, enabling data-driven control for IAB. The proposed architecture is illustrated in Fig.~\ref{fig:iab}. The main component is the extension of O-RAN interfaces (e.g., E2, O1, and O2) to the wireless domain of the network, through dedicated tunnels which allow the \glspl{ric} and \gls{smo} to reach the \gls{du} and \gls{mt} in the \gls{iab}-Nodes.

\vspace{-.15cm}
\subsection{Research Directions}

\textbf{Extension of Open and Intelligent IAB to NTN.} 6G networks will likely integrate components associated to \gls{ntn}, either for backhaul or also directly for access~\cite{9711564,giordani2021non}. How to extend O-RAN-based \gls{iab} and self-backhaul optimization to the \gls{ntn} domain is an open challenge~\cite{campana2023oran}. 
Here future research efforts intersect with the discussions on support for \gls{ntn} in the cellular architecture~\cite{abdullah2023integrated,saafi2022ai}, including (i) evaluations of different splits for the \glspl{gnb} and \gls{iab} systems (i.e., on whether the satellite is a physical layer relay or if it comes with the higher layers of the stack); (ii) mobility and handover management across the terrestrial and the \gls{ntn} domain; and (iii) network topology and architectures for the \gls{ntn} component, which could be served by a variety of non-terrestrial devices including satellites in different orbits, \glspl{uav}, baloons, among others. In this context, the intelligent control, the openness, and the softwarization brought by the Open RAN can enable dynamic optimization of the resources and of the topology tree, thus it will be important to manage the mobility and dynamics associated to the \gls{ntn} systems.

\textbf{Integration of IAB Nodes, Virtualization, and Dynamic Scaling.} One of the key advantages of the softwarized infrastructure discussed in Sec.~\ref{sec:new-arch} is the possibility of seamlessly updating, scaling, and powering on/off network functions according to the actual network needs. Dynamic scaling is a challenge when it comes to \gls{iab} nodes~\cite{gemmi2023globecom}, as turning off the radios in a self-backhauled device implies breaking the link and communications with the upstream and downstream nodes. The latter would also need to be reallocated to other upstream nodes, causing multiple handovers and reconnection attempts. Therefore, there is significant interest toward design of multi-radio \gls{iab} nodes, which exploit multi-connectivity with low-power radios or wake-up radios to enable dynamic scaling of resources in an \gls{iab} tree.

\section{Conclusions}
\label{sec:conclusions}

In this paper, we reviewed how the Open \gls{ran} paradigm is a key enabler of innovations in 6G networks, considering a system-level and architectural perspective. We have discussed key requirements for 6G, including energy efficiency, coverage, resiliency, cost and complexity, and innovation, and explored how Open \gls{ran} design principles connect to such 6G requirements. For each principle, we have presented what are the open challenges associated to its full development and deployment in commercial networks.

This tutorial has highlighted how openness, virtualization, programmability, softwarization, scaling, spectrum sharing, self-backhaul, optimization, and automation are core components of future \gls{6g} systems, and, in general, of how cellular network need to be deployed going forward. There are still several open challenges, primarily related to the design and testing of intelligent algorithms that can fully take advantage of such primitives to drive the network efficiency and performance, which we believe need to be the focus of the wireless networking research community as we head into \gls{6g}.

\footnotesize  % for natbib
\bibliographystyle{IEEEtran}
\bibliography{biblio.bib}

\begin{IEEEbiography}
[{\includegraphics[width=1in,height=1.25in,keepaspectratio]{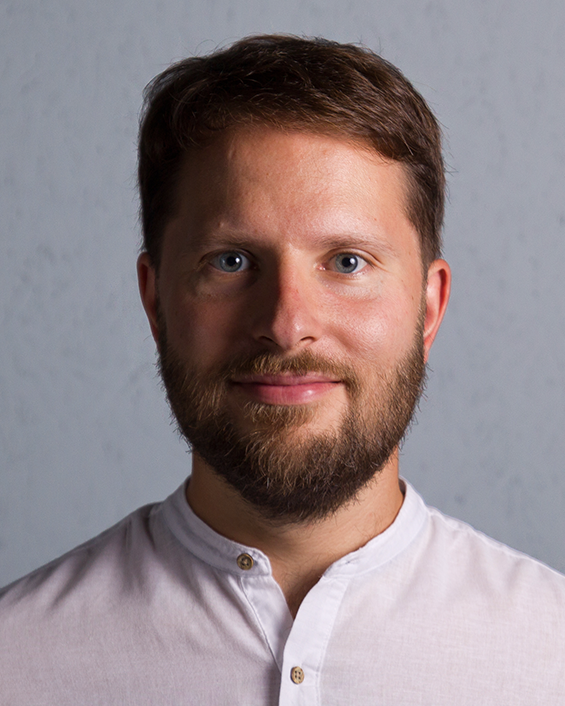}}]{Michele Polese} is a Research Assistant Professor at the Institute for the Wireless Internet of Things, Northeastern University, Boston, since October 2023. He received his Ph.D. at the Department of Information Engineering of the University of Padova in 2020. He then joined Northeastern University as a research scientist and part-time lecturer in 2020. During his Ph.D., he visited New York University (NYU), AT\&T Labs in Bedminster, NJ, and Northeastern University.
His research interests are in the analysis and development of protocols and architectures for future generations of cellular networks (5G and beyond), in particular for millimeter-wave and terahertz networks, spectrum sharing and passive/active user coexistence, open RAN development, and the performance evaluation of end-to-end, complex networks. He has contributed to O-RAN technical specifications and submitted responses to multiple FCC and NTIA notice of inquiry and requests for comments, and is a member of the Committee on Radio Frequency Allocations of the American Meteorological Society (2022-2024). He was awarded with several best paper awards, is serving as TPC co-chair for WNS3 2021-2022, as an Associate Technical Editor for the IEEE Communications Magazine, and has organized the Open 5G Forum in Fall 2021. He is a Member of the IEEE.
\end{IEEEbiography}

\begin{IEEEbiography}
[{\includegraphics[width=1in,height=1.25in,keepaspectratio]{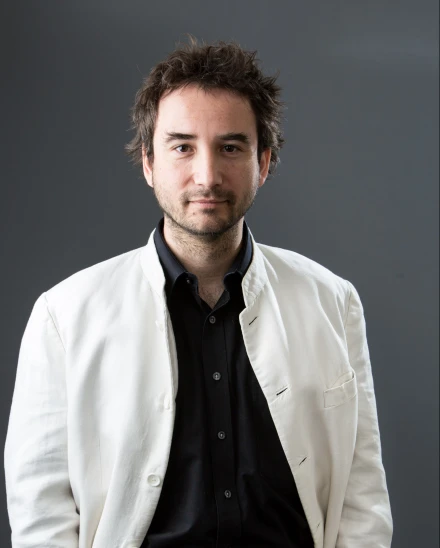}}]
{Mischa Dohler} Mischa Dohler is now VP Emerging Technologies at Ericsson Inc. in Silicon Valley, working on cutting-edge topics of 6G, Metaverse, XR, Quantum and Blockchain. He serves on the Technical Advisory Committee of the FCC and on the Spectrum Advisory Board of Ofcom.
He is a Fellow of the IEEE, the Royal Academy of Engineering, the Royal Society of Arts (RSA), the Institution of Engineering and Technology (IET); and a Distinguished Member of Harvard Square Leaders Excellence. He is a serial entrepreneur with 5 companies; composer \& pianist with 5 albums on Spotify/iTunes; and fluent in several languages. He has had ample coverage by national and international press and media, and is featured on Amazon Prime.

He is a frequent keynote, panel and tutorial speaker, and has received numerous awards. He has pioneered several research fields, contributed to numerous wireless broadband, IoT/M2M and cyber security standards, holds a dozen patents, organized and chaired numerous conferences, was the Editor-in-Chief of two journals, has more than 300 highly-cited publications, and authored several books. He is a Top-1\% Cited Innovator across all science fields globally.
He was Professor in Wireless Communications at King’s College London and Director of the Centre for Telecommunications Research from 2013-2021, driving cross-disciplinary research and innovation in technology, sciences and arts. He is the Cofounder and former CTO of the IoT-pioneering company Worldsensing; cofounder and former CTO of the AI-driven satellite company SiriusInsight.AI, and cofounder of the sustainability company Movingbeans. He also worked as a Senior Researcher at Orange/France Telecom from 2005-2008.
\end{IEEEbiography}

\begin{IEEEbiography}
[{\includegraphics[width=1in,height=1.25in,keepaspectratio]{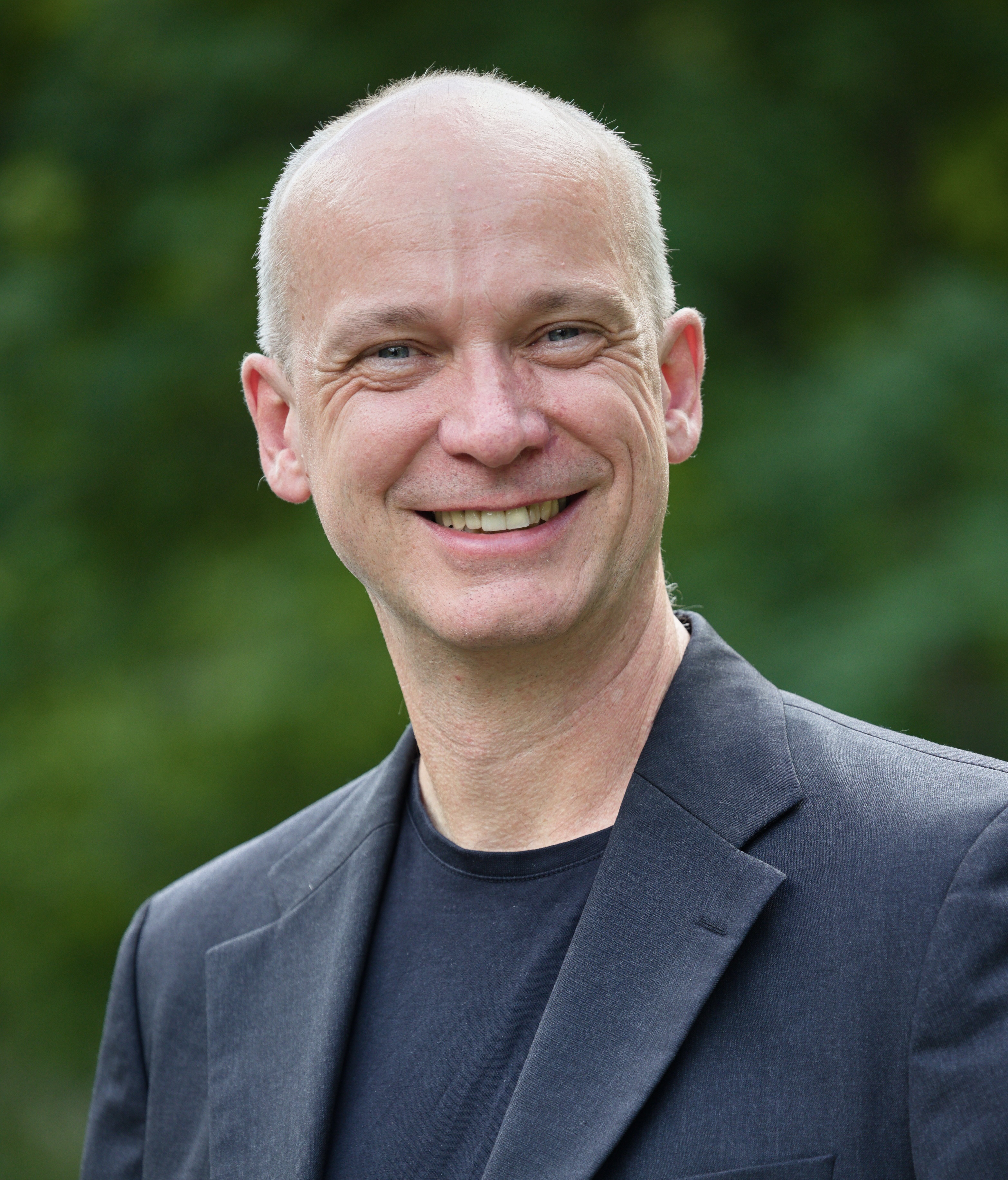}}]
{Falko Dressler} is full professor and Chair for Telecommunication Networks at the School of Electrical Engineering and Computer Science, TU Berlin. He received his M.Sc. and Ph.D. degrees from the Dept. of Computer Science, University of Erlangen in 1998 and 2003, respectively.
Dr. Dressler has been associate editor-in-chief for IEEE Trans. on Mobile Computing and Elsevier Computer Communications as well as an editor for journals such as IEEE/ACM Trans. on Networking, IEEE Trans. on Network Science and Engineering, Elsevier Ad Hoc Networks, and Elsevier Nano Communication Networks. He has been chairing conferences such as IEEE INFOCOM, ACM MobiSys, ACM MobiHoc, IEEE VNC, IEEE GLOBECOM. He authored the textbooks Self-Organization in Sensor and Actor Networks published by Wiley \& Sons and Vehicular Networking published by Cambridge University Press. He has been an IEEE Distinguished Lecturer as well as an ACM Distinguished Speaker.
Dr. Dressler is an IEEE Fellow as well as an ACM Distinguished Member. He is a member of the German National Academy of Science and Engineering (acatech). He has been serving on the IEEE COMSOC Conference Council and the ACM SIGMOBILE Executive Committee. His research objectives include adaptive wireless networking (sub-6GHz, mmWave, visible light, molecular communication) and wireless-based sensing with applications in ad hoc and sensor networks, the Internet of Things, and Cyber-Physical Systems.
\end{IEEEbiography}
\begin{IEEEbiography}
[{\includegraphics[width=1in,height=1.25in,keepaspectratio]{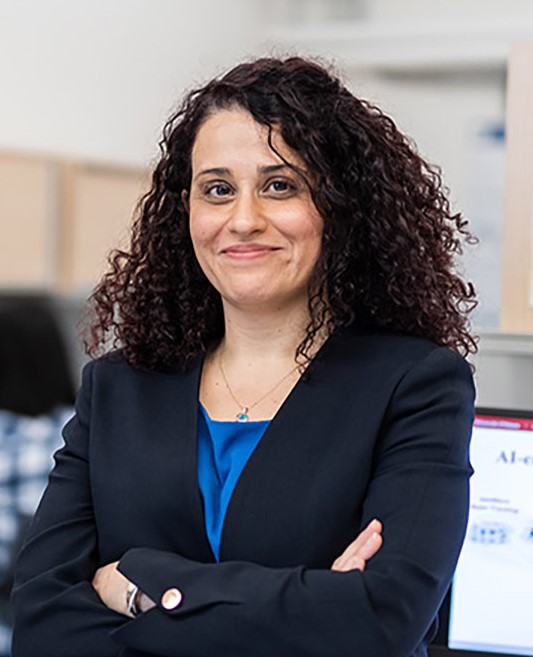}}]
{Melike Erol-Kantarci} is Chief Cloud RAN AI ML Data Scientist at Ericsson and Canada Research Chair in AI-enabled Next-Generation Wireless Networks and Full Professor at the School of Electrical Engineering and Computer Science at the University of Ottawa. She is the founding director of the Networked Systems and Communications Research (NETCORE) laboratory. She has received numerous awards and recognition. Dr. Erol-Kantarci is the co-editor of three books on smart grids, smart cities and intelligent transportation. She has over 200 peer-reviewed publications. She has delivered 70+ keynotes, plenary talks and tutorials around the globe. Dr. Erol-Kantarci is on the editorial board of the IEEE Transactions on Communications, IEEE Transactions on Cognitive Communications and Networking and IEEE Networking Letters.  She has acted as the general chair and technical program chair for many international conferences and workshops. Her main research interests are AI-enabled wireless networks, 5G and 6G wireless communications, smart grid and Internet of Things. Dr. Erol-Kantarci is an IEEE ComSoc Distinguished Lecturer, IEEE Senior member and ACM Senior Member.
\end{IEEEbiography}

\begin{IEEEbiography}
[{\includegraphics[width=1in,height=1.25in,keepaspectratio]{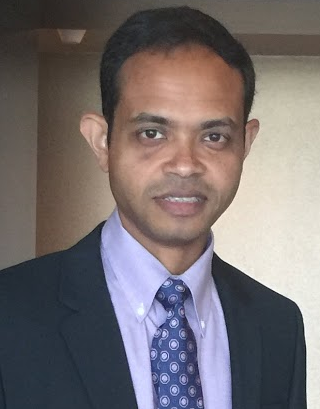}}]
{Rittwik Jana}
is currently a Principal Wireless Network Engineer at Google. Prior to this he worked at VMWare as Chief architect of RAN Intelligence and at AT\&T Labs Research. His interests span architecting the disaggregated RAN intelligent controller in O-RAN, service composition of VNFs using TOSCA, model driven control loop and automation in ONAP, video streaming for cellular networks and systems. He is a recipient of the AT\&T Science and Technology medal in 2016 for contributions to model driven cellular network planning, the IEEE Jack Neubauer memorial award in 2017 for systems work on full duplex wireless, and of several best paper awards in wireless communications. He holds 100 granted patents and has published over 150 technical papers at IEEE and ACM conferences and journals.
\end{IEEEbiography}

\begin{IEEEbiography}
[{\includegraphics[width=1in,height=1.25in,keepaspectratio]{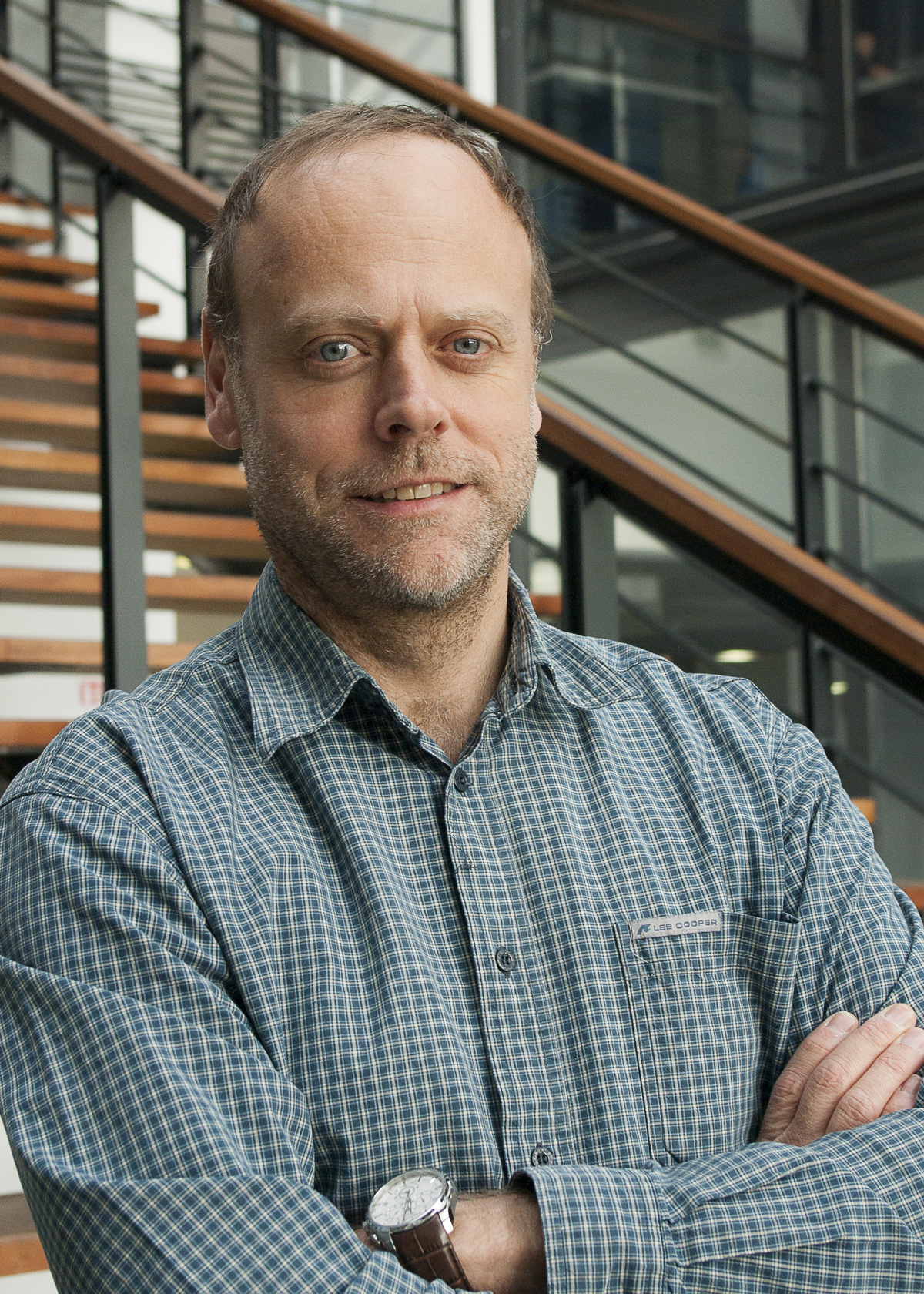}}]
{Raymond Knopp} received his PhD degree in Communication Systems from the Swiss Federal Institute of Technology (EPFL), Lausanne in 1997. 
His current research and teaching interests are in the areas of digital communications, software radio architectures, and implementation aspects of signal processing systems and real-time wireless networking protocols. He is currently a professor and the head of the Communication Systems Department at EURECOM, Sophia Antipolis, France. He is a leading figure in the OpenAirInterface (OAI) Community and has been instrumental in making open software for Radio Access Networks a reality through contributions over two decades. He is one of the very first and one of the most significant contributors to the OAI codebase. He was elected as the President of the OpenAirInterface Software Alliance (OSA), a non-profit organization federating the OAI and Mosaic5G codebases in December 2018.
\end{IEEEbiography}

\begin{IEEEbiography}
[{\includegraphics[width=1in,height=1.25in,keepaspectratio]{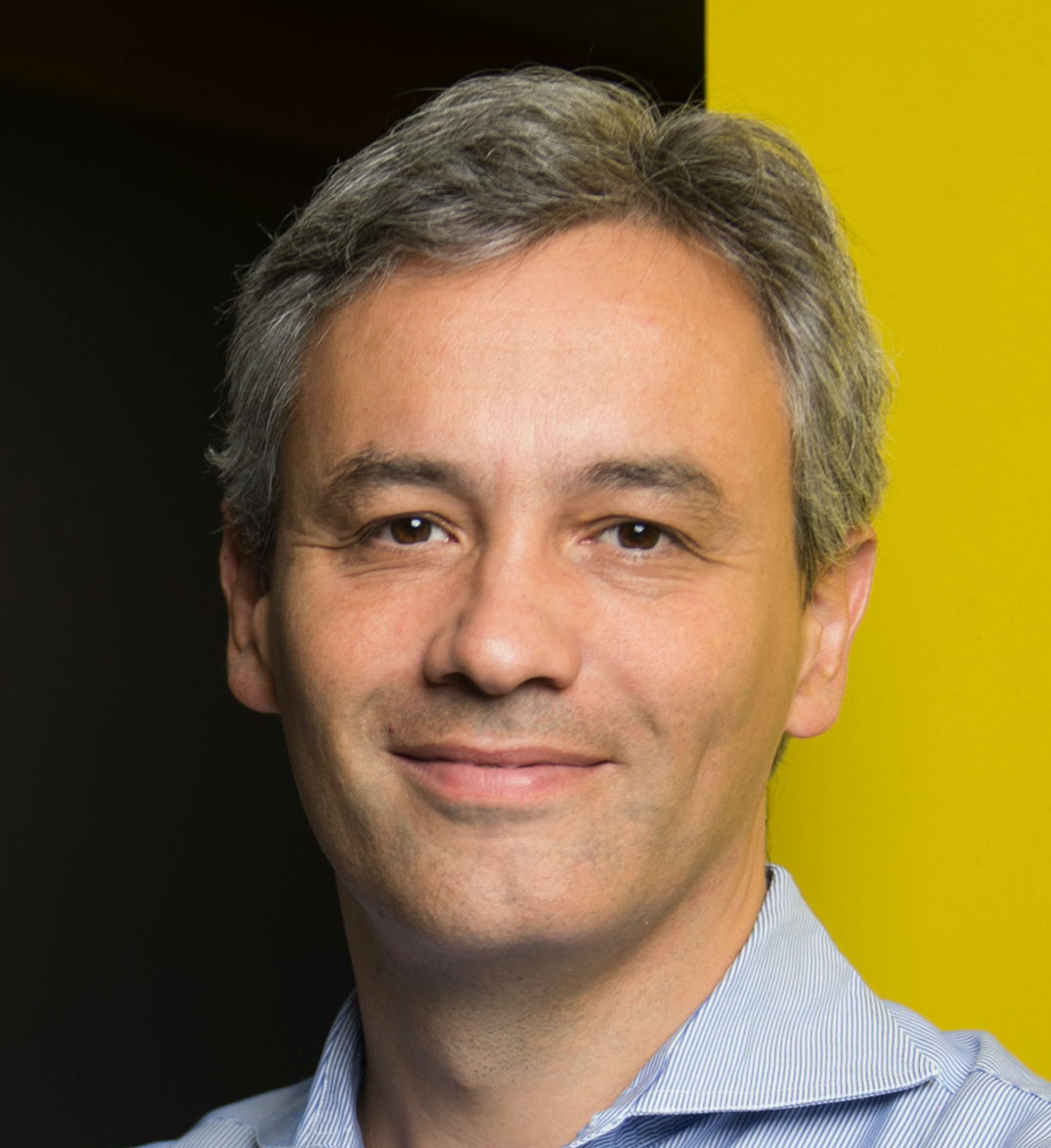}}]{Tommaso Melodia}
is the William Lincoln Smith Chair Professor with the Department of Electrical and Computer Engineering at Northeastern University in Boston. He is also the Founding Director of the Institute for the Wireless Internet of Things and the Director of Research for the PAWR Project Office. He received his Ph.D. in Electrical and Computer Engineering from the Georgia Institute of Technology in 2007. He is a recipient of the National Science Foundation CAREER award. Prof. Melodia has served as Associate Editor of IEEE Transactions on Wireless Communications, IEEE Transactions on Mobile Computing, Elsevier Computer Networks, among others. He has served as Technical Program Committee Chair for IEEE Infocom 2018, General Chair for IEEE SECON 2019, ACM Nanocom 2019, and ACM WUWnet 2014. Prof. Melodia is the Director of Research for the Platforms for Advanced Wireless Research (PAWR) Project Office, a \$100M public-private partnership to establish 4 city-scale platforms for wireless research to advance the US wireless ecosystem in years to come. Prof. Melodia's research on modeling, optimization, and experimental evaluation of Internet-of-Things and wireless networked systems has been funded by the National Science Foundation, the Air Force Research Laboratory the Office of Naval Research, DARPA, and the Army Research Laboratory. Prof. Melodia is a Fellow of the IEEE and a Senior Member of the ACM.
\end{IEEEbiography}

\end{document}

%% file: acronyms.tex
\newacronym{3gpp}{3GPP}{3rd Generation Partnership Project}
\newacronym{4g}{4G}{4th generation}
\newacronym{5g}{5G}{5th generation}
\newacronym{6g}{6G}{6th generation}
\newacronym{5gc}{5GC}{5G Core}
\newacronym{adc}{ADC}{Analog to Digital Converter}
\newacronym{aerpaw}{AERPAW}{Aerial Experimentation and Research Platform for Advanced Wireless}
\newacronym{ai}{AI}{Artificial Intelligence}
\newacronym{aimd}{AIMD}{Additive Increase Multiplicative Decrease}
\newacronym{am}{AM}{Acknowledged Mode}
\newacronym{amc}{AMC}{Adaptive Modulation and Coding}
\newacronym{amf}{AMF}{Access and Mobility Management Function}
\newacronym{aops}{AOPS}{Adaptive Order Prediction Scheduling}
\newacronym{api}{API}{Application Programming Interface}
\newacronym{apn}{APN}{Access Point Name}
\newacronym{ap}{AP}{Application Protocol}
\newacronym{aqm}{AQM}{Active Queue Management}
\newacronym{ar}{AR}{Augmented Reality}
\newacronym{ausf}{AUSF}{Authentication Server Function}
\newacronym{avc}{AVC}{Advanced Video Coding}
\newacronym{awgn}{AGWN}{Additive White Gaussian Noise}
\newacronym{balia}{BALIA}{Balanced Link Adaptation Algorithm}
\newacronym{bbu}{BBU}{Base Band Unit}
\newacronym{bdp}{BDP}{Bandwidth-Delay Product}
\newacronym{ber}{BER}{Bit Error Rate}
\newacronym{bf}{BF}{Beamforming}
\newacronym{bler}{BLER}{Block Error Rate}
\newacronym{brr}{BRR}{Bayesian Ridge Regressor}
\newacronym{bs}{BS}{Base Station}
\newacronym{bsr}{BSR}{Buffer Status Report}
\newacronym{bss}{BSS}{Business Support System}
\newacronym{ca}{CA}{Carrier Aggregation}
\newacronym{caas}{CaaS}{Connectivity-as-a-Service}
\newacronym{cb}{CB}{Code Block}
\newacronym{cc}{CC}{Congestion Control}
\newacronym{ccid}{CCID}{Congestion Control ID}
\newacronym{cco}{CC}{Carrier Component}
\newacronym{cdd}{CDD}{Cyclic Delay Diversity}
\newacronym{cdf}{CDF}{Cumulative Distribution Function}
\newacronym{cdn}{CDN}{Content Distribution Network}
\newacronym{cn}{CN}{core network}
\newacronym{codel}{CoDel}{Controlled Delay Management}
\newacronym{comac}{COMAC}{Converged Multi-Access and Core}
\newacronym{cord}{CORD}{Central Office Re-architected as a Datacenter}
\newacronym{cornet}{CORNET}{COgnitive Radio NETwork}
\newacronym{cosmos}{COSMOS}{Cloud Enhanced Open Software Defined Mobile Wireless Testbed for City-Scale Deployment}
\newacronym{cots}{COTS}{Commercial Off-the-Shelf}
\newacronym{cp}{CP}{Control Plane}
\newacronym{cyp}{CP}{Cyclic Prefix}
\newacronym{up}{UP}{User Plane}
\newacronym{cpu}{CPU}{Central Processing Unit}
\newacronym{cqi}{CQI}{Channel Quality Information}
\newacronym{cr}{CR}{Cognitive Radio}
\newacronym{cran}{C-RAN}{Cloud \gls{ran}}
\newacronym{crs}{CRS}{Cell Reference Signal}
\newacronym{csi}{CSI}{Channel State Information}
\newacronym{csirs}{CSI-RS}{Channel State Information - Reference Signal}
\newacronym{cu}{CU}{Central Unit}
\newacronym{d2tcp}{D$^2$TCP}{Deadline-aware Data center TCP}
\newacronym{d3}{D$^3$}{Deadline-Driven Delivery}
\newacronym{dac}{DAC}{Digital to Analog Converter}
\newacronym{dag}{DAG}{Directed Acyclic Graph}
\newacronym{das}{DAS}{Distributed Antenna System}
\newacronym{dash}{DASH}{Dynamic Adaptive Streaming over HTTP}
\newacronym{dc}{DC}{Dual Connectivity}
\newacronym{dccp}{DCCP}{Datagram Congestion Control Protocol}
\newacronym{dce}{DCE}{Direct Code Execution}
\newacronym{dci}{DCI}{Downlink Control Information}
\newacronym{dctcp}{DCTCP}{Data Center TCP}
\newacronym{dl}{DL}{Downlink}
\newacronym{dmr}{DMR}{Deadline Miss Ratio}
\newacronym{dmrs}{DMRS}{DeModulation Reference Signal}
\newacronym{drlcc}{DRL-CC}{Deep Reinforcement Learning Congestion Control}
\newacronym{drs}{DRS}{Discovery Reference Signal}
\newacronym{du}{DU}{Distributed Unit}
\newacronym{e2e}{E2E}{end-to-end}
\newacronym{ecaas}{ECaaS}{Edge-Cloud-as-a-Service}
\newacronym{ecn}{ECN}{Explicit Congestion Notification}
\newacronym{edf}{EDF}{Earliest Deadline First}
\newacronym{embb}{eMBB}{Enhanced Mobile Broadband}
\newacronym{empower}{EMPOWER}{EMpowering transatlantic PlatfOrms for advanced WirEless Research}
\newacronym{enb}{eNB}{evolved Node Base}
\newacronym{endc}{EN-DC}{E-UTRAN-\gls{nr} \gls{dc}}
\newacronym{epc}{EPC}{Evolved Packet Core}
\newacronym{eps}{EPS}{Evolved Packet System}
\newacronym{es}{ES}{Edge Server}
\newacronym{etsi}{ETSI}{European Telecommunications Standards Institute}
\newacronym[firstplural=Estimated Times of Arrival (ETAs)]{eta}{ETA}{Estimated Time of Arrival}
\newacronym{eutran}{E-UTRAN}{Evolved Universal Terrestrial Access Network}
\newacronym{faas}{FaaS}{Function-as-a-Service}
\newacronym{fapi}{FAPI}{Functional Application Platform Interface}
\newacronym{fdd}{FDD}{Frequency Division Duplexing}
\newacronym{fdm}{FDM}{Frequency Division Multiplexing}
\newacronym{fdma}{FDMA}{Frequency Division Multiple Access}
\newacronym{fed4fire}{FED4FIRE+}{Federation 4 Future Internet Research and Experimentation Plus}
\newacronym{fir}{FIR}{Finite Impulse Response}
\newacronym{fit}{FIT}{Future \acrlong{iot}}
\newacronym{fpga}{FPGA}{Field Programmable Gate Array}
\newacronym{fr2}{FR2}{Frequency Range 2}
\newacronym{fs}{FS}{Fast Switching}
\newacronym{fscc}{FSCC}{Flow Sharing Congestion Control}
\newacronym{ftp}{FTP}{File Transfer Protocol}
\newacronym{fw}{FW}{Flow Window}
\newacronym{ge}{GE}{Gaussian Elimination}
\newacronym{gnb}{gNB}{Next Generation Node Base}
\newacronym{gop}{GOP}{Group of Pictures}
\newacronym{gpr}{GPR}{Gaussian Process Regressor}
\newacronym{gpu}{GPU}{Graphics Processing Unit}
\newacronym{gtp}{GTP}{GPRS Tunneling Protocol}
\newacronym{gtpc}{GTP-C}{GPRS Tunnelling Protocol Control Plane}
\newacronym{gtpu}{GTP-U}{GPRS Tunnelling Protocol User Plane}
\newacronym{gtpv2c}{GTPv2-C}{\gls{gtp} v2 - Control}
\newacronym{gw}{GW}{Gateway}
\newacronym{harq}{HARQ}{Hybrid Automatic Repeat reQuest}
\newacronym{hetnet}{HetNet}{Heterogeneous Network}
\newacronym{hh}{HH}{Hard Handover}
\newacronym{hol}{HOL}{Head-of-Line}
\newacronym{hqf}{HQF}{Highest-quality-first}
\newacronym{hss}{HSS}{Home Subscription Server}
\newacronym{http}{HTTP}{HyperText Transfer Protocol}
\newacronym{ia}{IA}{Initial Access}
\newacronym{iab}{IAB}{Integrated Access and Backhaul}
\newacronym{ic}{IC}{Incident Command}
\newacronym{ietf}{IETF}{Internet Engineering Task Force}
\newacronym{imsi}{IMSI}{International Mobile Subscriber Identity}
\newacronym{imt}{IMT}{International Mobile Telecommunication}
\newacronym{iot}{IoT}{Internet of Things}
\newacronym{ip}{IP}{Internet Protocol}
\newacronym{itu}{ITU}{International Telecommunication Union}
\newacronym{kpi}{KPI}{Key Performance Indicator}
\newacronym{kpm}{KPM}{Key Performance Measurement}
\newacronym{kvm}{KVM}{Kernel-based Virtual Machine}
\newacronym{los}{LOS}{Line-of-Sight}
\newacronym{lsm}{LSM}{Link-to-System Mapping}
\newacronym{lstm}{LSTM}{Long Short Term Memory}
\newacronym{lte}{LTE}{Long Term Evolution}
\newacronym{lxc}{LXC}{Linux Container}
\newacronym{m2m}{M2M}{Machine to Machine}
\newacronym{mac}{MAC}{Medium Access Control}
\newacronym{manet}{MANET}{Mobile Ad Hoc Network}
\newacronym{mano}{MANO}{Management and Orchestration}
\newacronym{mc}{MC}{Multi-Connectivity}
\newacronym{mcc}{MCC}{Mobile Cloud Computing}
\newacronym{mchem}{MCHEM}{Massive Channel Emulator}
\newacronym{mcs}{MCS}{Modulation and Coding Scheme}
\newacronym{mec}{MEC}{Multi-access Edge Computing}
\newacronym{mec2}{MEC}{Mobile Edge Cloud}
\newacronym{mfc}{MFC}{Mobile Fog Computing}
\newacronym{mgen}{MGEN}{Multi-Generator}
\newacronym{mi}{MI}{Mutual Information}
\newacronym{mib}{MIB}{Master Information Block}
\newacronym{miesm}{MIESM}{Mutual Information Based Effective SINR}
\newacronym{mimo}{MIMO}{Multiple Input, Multiple Output}
\newacronym{ml}{ML}{Machine Learning}
\newacronym{mlr}{MLR}{Maximum-local-rate}
\newacronym[plural=\gls{mme}s,firstplural=Mobility Management Entities (MMEs)]{mme}{MME}{Mobility Management Entity}
\newacronym{mmtc}{mMTC}{Massive Machine-Type Communications}
\newacronym{mmwave}{mmWave}{millimeter wave}
\newacronym{mpdccp}{MP-DCCP}{Multipath Datagram Congestion Control Protocol}
\newacronym{mptcp}{MPTCP}{Multipath TCP}
\newacronym{mr}{MR}{Maximum Rate}
\newacronym{mrdc}{MR-DC}{Multi \gls{rat} \gls{dc}}
\newacronym{mse}{MSE}{Mean Square Error}
\newacronym{mss}{MSS}{Maximum Segment Size}
\newacronym{mt}{MT}{Mobile Termination}
\newacronym{mtd}{MTD}{Machine-Type Device}
\newacronym{mtu}{MTU}{Maximum Transmission Unit}
\newacronym{mumimo}{MU-MIMO}{Multi-user \gls{mimo}}
\newacronym{mvno}{MVNO}{Mobile Virtual Network Operator}
\newacronym{nalu}{NALU}{Network Abstraction Layer Unit}
\newacronym{nas}{NAS}{Non-Access Stratum}
\newacronym{nbiot}{NB-IoT}{Narrow Band IoT}
\newacronym{nfv}{NFV}{network function virtualization}
\newacronym{nfvi}{NFVI}{Network Function Virtualization Infrastructure}
\newacronym{ngrg}{nGRG}{next Generation Research Group}
\newacronym{ni}{NI}{Network Interfaces}
\newacronym{nic}{NIC}{Network Interface Card}
\newacronym{nlos}{NLOS}{Non-Line-of-Sight}
\newacronym{now}{NOW}{Non Overlapping Window}
\newacronym{nsm}{NSM}{Network Service Mesh}
\newacronym{nr}{NR}{New Radio}
\newacronym{nrf}{NRF}{Network Repository Function}
\newacronym{nsa}{NSA}{Non Stand Alone}
\newacronym{nse}{NSE}{Network Slicing Engine}
\newacronym{nssf}{NSSF}{Network Slice Selection Function}
\newacronym{o2i}{O2I}{Outdoor to Indoor}
\newacronym{oai}{OAI}{OpenAirInterface}
\newacronym{oaicn}{OAI-CN}{\gls{oai} \acrlong{cn}}
\newacronym{oairan}{OAI-RAN}{\acrlong{oai} \acrlong{ran}}
\newacronym{oam}{OAM}{Operations, Administration and Maintenance}
\newacronym{ofdm}{OFDM}{Orthogonal Frequency Division Multiplexing}
\newacronym{olia}{OLIA}{Opportunistic Linked Increase Algorithm}
\newacronym{omec}{OMEC}{Open Mobile Evolved Core}
\newacronym{onap}{ONAP}{Open Network Automation Platform}
\newacronym{onf}{ONF}{Open Networking Foundation}
\newacronym{onos}{ONOS}{Open Networking Operating System}
\newacronym{oom}{OOM}{\gls{onap} Operations Manager}
\newacronym{opnfv}{OPNFV}{Open Platform for \gls{nfv}}
\newacronym{oran}{O-RAN}{Open \gls{ran}}
\newacronym{orbit}{ORBIT}{Open-Access Research Testbed for Next-Generation Wireless Networks}
\newacronym{os}{OS}{Operating System}
\newacronym{oss}{OSS}{Operations Support System}
\newacronym{otic}{OTIC}{Open Testing \& Integration Centre}
\newacronym{pa}{PA}{Position-aware}
\newacronym{pase}{PASE}{Prioritization, Arbitration, and Self-adjusting Endpoints}
\newacronym{pawr}{PAWR}{Platforms for Advanced Wireless Research}
\newacronym{pbch}{PBCH}{Physical Broadcast Channel}
\newacronym{pcef}{PCEF}{Policy and Charging Enforcement Function}
\newacronym{pcfich}{PCFICH}{Physical Control Format Indicator Channel}
\newacronym{pcrf}{PCRF}{Policy and Charging Rules Function}
\newacronym{pdcch}{PDCCH}{Physical Downlink Control Channel}
\newacronym{pdcp}{PDCP}{Packet Data Convergence Protocol}
\newacronym{pdsch}{PDSCH}{Physical Downlink Shared Channel}
\newacronym{pdu}{PDU}{Packet Data Unit}
\newacronym{pf}{PF}{Proportional Fair}
\newacronym{pgw}{PGW}{Packet Gateway}
\newacronym{phich}{PHICH}{Physical Hybrid ARQ Indicator Channel}
\newacronym{phy}{PHY}{Physical}
\newacronym{pmch}{PMCH}{Physical Multicast Channel}
\newacronym{pmi}{PMI}{Precoding Matrix Indicators}
\newacronym{powder}{POWDER}{Platform for Open Wireless Data-driven Experimental Research}
\newacronym{ppo}{PPO}{Proximal Policy Optimization}
\newacronym{ppp}{PPP}{Poisson Point Process}
\newacronym{prach}{PRACH}{Physical Random Access Channel}
\newacronym{prb}{PRB}{Physical Resource Block}
\newacronym{psnr}{PSNR}{Peak Signal to Noise Ratio}
\newacronym{pss}{PSS}{Primary Synchronization Signal}
\newacronym{pucch}{PUCCH}{Physical Uplink Control Channel}
\newacronym{pusch}{PUSCH}{Physical Uplink Shared Channel}
\newacronym{qam}{QAM}{Quadrature Amplitude Modulation}
\newacronym{qci}{QCI}{\gls{qos} Class Identifier}
\newacronym{qoe}{QoE}{Quality of Experience}
\newacronym{qos}{QoS}{Quality of Service}
\newacronym{quic}{QUIC}{Quick UDP Internet Connections}
\newacronym{rach}{RACH}{Random Access Channel}
\newacronym{ran}{RAN}{radio access network}
\newacronym[firstplural=Radio Access Technologies (RATs)]{rat}{RAT}{Radio Access Technology}
\newacronym{rcn}{RCN}{Research Coordination Network}
\newacronym{rc}{RC}{RAN Control}
\newacronym{rec}{REC}{Radio Edge Cloud}
\newacronym{red}{RED}{Random Early Detection}
\newacronym{renew}{RENEW}{Reconfigurable Eco-system for Next-generation End-to-end Wireless}
\newacronym{rf}{RF}{Radio Frequency}
\newacronym{rfc}{RFC}{Request for Comments}
\newacronym{rfr}{RFR}{Random Forest Regressor}
\newacronym{ric}{RIC}{\gls{ran} Intelligent Controller}
\newacronym{rlc}{RLC}{Radio Link Control}
\newacronym{rlf}{RLF}{Radio Link Failure}
\newacronym{rlnc}{RLNC}{Random Linear Network Coding}
\newacronym{rmr}{RMR}{RIC Message Router}
\newacronym{rmse}{RMSE}{Root Mean Squared Error}
\newacronym{rnis}{RNIS}{Radio Network Information Service}
\newacronym{rr}{RR}{Round Robin}
\newacronym{rrc}{RRC}{Radio Resource Control}
\newacronym{rrm}{RRM}{Radio Resource Management}
\newacronym{rru}{RRU}{Remote Radio Unit}
\newacronym{rs}{RS}{Remote Server}
\newacronym{rsrp}{RSRP}{Reference Signal Received Power}
\newacronym{rsrq}{RSRQ}{Reference Signal Received Quality}
\newacronym{rss}{RSS}{Received Signal Strength}
\newacronym{rssi}{RSSI}{Received Signal Strength Indicator}
\newacronym{rtt}{RTT}{Round Trip Time}
\newacronym{ru}{RU}{Radio Unit}
\newacronym{rw}{RW}{Receive Window}
\newacronym{rx}{RX}{Receiver}
\newacronym{s1ap}{S1AP}{S1 Application Protocol}
\newacronym{sa}{SA}{standalone}
\newacronym{sack}{SACK}{Selective Acknowledgment}
\newacronym{sap}{SAP}{Service Access Point}
\newacronym{sc2}{SC2}{Spectrum Collaboration Challenge}
\newacronym{scef}{SCEF}{Service Capability Exposure Function}
\newacronym{sch}{SCH}{Secondary Cell Handover}
\newacronym{scoot}{SCOOT}{Split Cycle Offset Optimization Technique}
\newacronym{sctp}{SCTP}{Stream Control Transmission Protocol}
\newacronym{sdap}{SDAP}{Service Data Adaptation Protocol}
\newacronym{sdk}{SDK}{Software Development Kit}
\newacronym{sdm}{SDM}{Space Division Multiplexing}
\newacronym{sdma}{SDMA}{Spatial Division Multiple Access}
\newacronym{sdn}{SDN}{Software-defined Networking}
\newacronym{sdr}{SDR}{Software-defined Radio}
\newacronym{seba}{SEBA}{SDN-Enabled Broadband Access}
\newacronym{sgsn}{SGSN}{Serving GPRS Support Node}
\newacronym{sgw}{SGW}{Service Gateway}
\newacronym{si}{SI}{Study Item}
\newacronym{sib}{SIB}{Secondary Information Block}
\newacronym{sinr}{SINR}{Signal to Interference plus Noise Ratio}
\newacronym{sip}{SIP}{Session Initiation Protocol}
\newacronym{siso}{SISO}{Single Input, Single Output}
\newacronym{sla}{SLA}{Service Level Agreement}
\newacronym{sm}{SM}{Service Model}
\newacronym{smf}{SMF}{Session Management Function}
\newacronym{smo}{SMO}{Service Management and Orchestration}
\newacronym{sms}{SMS}{Short Message Service}
\newacronym{smsgmsc}{SMS-GMSC}{\gls{sms}-Gateway}
\newacronym{snr}{SNR}{Signal-to-Noise-Ratio}
\newacronym{son}{SON}{Self-Organizing Network}
\newacronym{sptcp}{SPTCP}{Single Path TCP}
\newacronym{srb}{SRB}{Service Radio Bearer}
\newacronym{srn}{SRN}{Standard Radio Node}
\newacronym{srs}{SRS}{Sounding Reference Signal}
\newacronym{ss}{SS}{Synchronization Signal}
\newacronym{sss}{SSS}{Secondary Synchronization Signal}
\newacronym{st}{ST}{Spanning Tree}
\newacronym{svc}{SVC}{Scalable Video Coding}
\newacronym{tb}{TB}{Transport Block}
\newacronym{tcp}{TCP}{Transmission Control Protocol}
\newacronym{tdd}{TDD}{Time Division Duplexing}
\newacronym{tdm}{TDM}{Time Division Multiplexing}
\newacronym{tdma}{TDMA}{Time Division Multiple Access}
\newacronym{tfl}{TfL}{Transport for London}
\newacronym{tfrc}{TFRC}{TCP-Friendly Rate Control}
\newacronym{tft}{TFT}{Traffic Flow Template}
\newacronym{tgen}{TGEN}{Traffic Generator}
\newacronym{tip}{TIP}{Telecom Infra Project}
\newacronym{tm}{TM}{Transparent Mode}
\newacronym{to}{TO}{Telco Operator}
\newacronym{tr}{TR}{Technical Report}
\newacronym{trp}{TRP}{Transmitter Receiver Pair}
\newacronym{ts}{TS}{Technical Specification}
\newacronym{tti}{TTI}{Transmission Time Interval}
\newacronym{ttt}{TTT}{Time-to-Trigger}
\newacronym{tx}{TX}{Transmitter}
\newacronym{uas}{UAS}{Unmanned Aerial System}
\newacronym{uav}{UAV}{Unmanned Aerial Vehicle}
\newacronym{udm}{UDM}{Unified Data Management}
\newacronym{udp}{UDP}{User Datagram Protocol}
\newacronym{udr}{UDR}{Unified Data Repository}
\newacronym{ue}{UE}{User Equipment}
\newacronym{uhd}{UHD}{\gls{usrp} Hardware Driver}
\newacronym{ul}{UL}{Uplink}
\newacronym{um}{UM}{Unacknowledged Mode}
\newacronym{uml}{UML}{Unified Modeling Language}
\newacronym{upa}{UPA}{Uniform Planar Array}
\newacronym{upf}{UPF}{User Plane Function}
\newacronym{urllc}{URLLC}{Ultra Reliable and Low Latency Communications}
\newacronym{usa}{U.S.}{United States}
\newacronym{usim}{USIM}{Universal Subscriber Identity Module}
\newacronym{usrp}{USRP}{Universal Software Radio Peripheral}
\newacronym{utc}{UTC}{Urban Traffic Control}
\newacronym{vim}{VIM}{Virtualization Infrastructure Manager}
\newacronym{vm}{VM}{Virtual Machine}
\newacronym{vnf}{VNF}{Virtual Network Function}
\newacronym{volte}{VoLTE}{Voice over \gls{lte}}
\newacronym{voltha}{VOLTHA}{Virtual OLT HArdware Abstraction}
\newacronym{vr}{VR}{Virtual Reality}
\newacronym{vran}{vRAN}{Virtualized \gls{ran}}
\newacronym{vss}{VSS}{Video Streaming Server}
\newacronym{wbf}{WBF}{Wired Bias Function}
\newacronym{wf}{WF}{Waterfilling}
\newacronym{wg}{WG}{Working Group}
\newacronym{wlan}{WLAN}{Wireless Local Area Network}
\newacronym{osm}{OSM}{Open Source Management and Orchestration}
\newacronym{pnf}{PNF}{Physical Network Function}
\newacronym{drl}{DRL}{Deep Reinforcement Learning}
\newacronym{mtc}{MTC}{Machine-type Communications}
\newacronym{osc}{OSC}{O-RAN Software Community}
\newacronym{mns}{MnS}{Management Services}
\newacronym{ves}{VES}{\gls{vnf} Event Stream}
\newacronym{ei}{EI}{Enrichment Information}
\newacronym{fh}{FH}{Fronthaul}
\newacronym{fft}{FFT}{Fast Fourier Transform}
\newacronym{laa}{LAA}{Licensed-Assisted Access}
\newacronym{plfs}{PLFS}{Physical Layer Frequency Signals}
\newacronym{ptp}{PTP}{Precision Time Protocol}
\newacronym{asic}{ASIC}{application-specific integrated circuit}
\newacronym{aal}{AAL}{Acceleration Abstraction Layer}
\newacronym{fec}{FEC}{Forward Error Correction}
\newacronym{sdl}{SDL}{Shared Data Layer}
\newacronym{nib}{NIB}{Network Information Base}
\newacronym{rnib}{R-NIB}{RAN \gls{nib}}
\newacronym{fcaps}{FCAPS}{Fault, Configuration, Accounting, Performance, Security}
\newacronym{ie}{IE}{Information Element}
\newacronym{fg}{FG}{Focus Group}
\newacronym{osfg}{OSFG}{Open Source Focus Group}
\newacronym{sdfg}{SDFG}{Standard Development Focus Group}
\newacronym{tifg}{TIFG}{Test \& Integration Focus Group}
\newacronym{sfg}{SFG}{Security Focus Group}
\newacronym{swg}{SWG}{Security Work Group}
\newacronym{e2sm}{E2SM}{E2 Service Model}
\newacronym{tsc}{TSC}{Technical Steering Committee}
\newacronym{sdo}{SDO}{Standard-Development Organization}
\newacronym{sql}{SQL}{Structured Query Language}
\newacronym{ssh}{SSH}{Secure Shell}
\newacronym{tls}{TLS}{Transport Layer Security}
\newacronym{netconf}{NETCONF}{Network Configuration Protocol}
\newacronym{dtls}{DTLS}{Datagram Transport Layer Security}
\newacronym{cmp}{CMP}{Certificate Management Protocol}
\newacronym{ccc}{CCC}{Cell Configuration and Control}
\newacronym{dsp}{DSP}{Digital Signal Processing}
\newacronym{opex}{OPEX}{Operational Expenses}
\newacronym{cbrs}{CBRS}{Citizen Broadband Radio Service}
\newacronym{ntn}{NTN}{Non-terrestrial Network}

\newacronym{appmgr}{APPMGR}{App Manager}
\newacronym{near-rt}{Near-RT}{Near-real-time}
\newacronym{non-rt}{Non-RT}{Non-real-time}
\newacronym{sba}{SBA}{Service-based Architecture}
\newacronym{hrl}{HRL}{Hierarchical Reinforcement Learning}

\newacronym{onr}{ONR}{Office of Naval Research}
\newacronym{afosr}{AFOSR}{Air Force Office of Scientific Research}
\newacronym{afrl}{AFRL}{Air Force Research Laboratory}
\newacronym{arl}{ARL}{Army Research Laboratory}

\newacronym{arc}{ARC}{Aerial Research Cloud}

\newacronym{mno}{MNO}{Mobile Network Operator}
\newacronym{ct}{CT}{Continuous Testing}
\newacronym{oci}{OCI}{Open Container Initiative}
\newacronym{ci}{CI}{Continuous Integration}
\newacronym{cd}{CD}{Continuous Deployment}

%% file: myTikz.tex
\tikzstyle{startstop} = [rectangle, rounded corners, minimum width=2cm, minimum height=0.5cm,text centered, draw=black]
\tikzstyle{io} = [trapezium, trapezium left angle=70, trapezium right angle=110, minimum width=3cm, minimum height=1cm, text centered, draw=black]
\tikzstyle{process} = [rectangle, minimum width=2cm, minimum height=0.5cm, text centered, draw=black, alignb=center]
\tikzstyle{decision} = [ellipse, minimum width=2cm, minimum height=1cm, text centered, draw=black]
\tikzstyle{arrow} = [thick,<->,>=stealth]
\tikzstyle{line} = [thick,>=stealth]
\tikzstyle{darrow} = [thick,<->,>=stealth,dashed]
\tikzstyle{sarrow} = [thick,->,>=stealth]
\tikzstyle{larrow} = [line width=0.1mm,dashdotted,->,>=stealth]
\tikzstyle{llarrow} = [line width=0.1mm,->,>=stealth]

\makeatletter
\def\grd@save@target#1{%
  \def\grd@target{#1}}
\def\grd@save@start#1{%
  \def\grd@start{#1}}
\tikzset{
  grid with coordinates/.style={
    to path={%
      \pgfextra{%
        \edef\grd@@target{(\tikztotarget)}%
        \tikz@scan@one@point\grd@save@target\grd@@target\relax
        \edef\grd@@start{(\tikztostart)}%
        \tikz@scan@one@point\grd@save@start\grd@@start\relax
        \draw[minor help lines] (\tikztostart) grid (\tikztotarget);
        \draw[major help lines] (\tikztostart) grid (\tikztotarget);
        \grd@start
        \pgfmathsetmacro{\grd@xa}{\the\pgf@x/1cm}
        \pgfmathsetmacro{\grd@ya}{\the\pgf@y/1cm}
        \grd@target
        \pgfmathsetmacro{\grd@xb}{\the\pgf@x/1cm}
        \pgfmathsetmacro{\grd@yb}{\the\pgf@y/1cm}
        \pgfmathsetmacro{\grd@xc}{\grd@xa + \pgfkeysvalueof{/tikz/grid with coordinates/major step x}}
        \pgfmathsetmacro{\grd@yc}{\grd@ya + \pgfkeysvalueof{/tikz/grid with coordinates/major step y}}
        \foreach \x in {\grd@xa,\grd@xc,...,\grd@xb}
        \node[anchor=north] at (\x,\grd@ya) {\pgfmathprintnumber{\x}};
        \foreach \y in {\grd@ya,\grd@yc,...,\grd@yb}
        \node[anchor=east] at (\grd@xa,\y) {\pgfmathprintnumber{\y}};
      }
    }
  },
  minor help lines/.style={
    help lines,
    gray,
    line cap =round,
    xstep=\pgfkeysvalueof{/tikz/grid with coordinates/minor step x},
    ystep=\pgfkeysvalueof{/tikz/grid with coordinates/minor step y}
  },
  major help lines/.style={
    help lines,
    line cap =round,
    line width=\pgfkeysvalueof{/tikz/grid with coordinates/major line width},
    xstep=\pgfkeysvalueof{/tikz/grid with coordinates/major step x},
    ystep=\pgfkeysvalueof{/tikz/grid with coordinates/major step y}
  },
  grid with coordinates/.cd,
  minor step x/.initial=.5,
  minor step y/.initial=.2,
  major step x/.initial=1,
  major step y/.initial=1,
  major line width/.initial=1pt,
}
\makeatother